\documentclass[showpacs,preprintnumbers,amsmath,amssymb,aps, prd]{revtex4}
\usepackage{graphicx}% Include figure files
\usepackage{grffile}
\usepackage{caption}
\usepackage{subcaption}
\usepackage{float}
\usepackage{color}
\usepackage{dcolumn}% Align table columns on decimal point
\usepackage{bm}% bold math
\begin{document}
\title{Kerr-Sen-like Lorentz violating black holes and Superradiance phenomena}
% Force line breaks with \
\author{Sohan Kumar Jha}
\affiliation{Chandernagore College, Chandernagore, Hooghly, West
Bengal, India}
\author{Anisur Rahaman}
 \email{anisur.associates@aucaa.ac.in;
 manisurn@gmail.com (Corresponding Author)}
\affiliation{Durgapur Government College, Durgapur, Burdwan -
713214, West Bengal, India}

\date{\today}% It is always \today, today, %  but any date may be explicitly specified
\begin{abstract}
\begin{center}
Abstract
\end{center}
A Kerr-Sen-like black hole solution results from
Einstein-bumblebee gravity. It contains a Lorentz violating (LV)
parameter that enters when the bumblebee field receives vacuum
expectation value through a spontaneously breaking of the symmetry
of the classical action. The geometrical structure concerning the
singularity of this spacetime is studied with reference to the
parameters involved in the Kerr-Sen-like metric. We introduce this
Einstein-bumblebee modified gravity to probe the role of
spontaneous Lorentz violation on the superradiance scattering
phenomena and the instability associated with it. We observe that
for the low-frequency range of the scalar wave the superradiance
scattering gets enhanced when the Lorentz-violating parameter
$\ell$ takes the negative values and it reduces when values of
$\ell$ are positive. The study of the black hole bomb issue
reveals that for the negative values of $\ell$, the parameter
space of the scalar field instability increase prominently,
however, for its positive values, it shows a considerable
reduction. We also tried to put constraints on the parameters
contained in the Kerr-Sen-like black hole by comparing the
deformation of the shadow produced by the black hole parameters
with the observed deviation from circularity and the
angular deviation from the $M87*$ data.

\end{abstract}
\maketitle
\section{Introduction}
In a gravitational system, the scattering of radiation off
absorbing rotating objects produce waves with amplitude larger
than incident one under certain conditions which is known as
rotational superradiance \cite{ZEL0, ZEL1}. In 1971, Zel'dovich
showed that scattering of radiation off rotating absorbing
surfaces result in waves with a larger amplitude as $\omega <
m\Omega$ where $\omega $ is the frequency of the incident
monochromatic radiation with m, the azimuthal number with respect
to the rotation axis and $\Omega$ is the angular velocity of the
rotating gravitational system. For review we would like to mention
the lecture notes \cite{REVIEW}, and the references therein.
Rotational superradiance belongs to a wider class of classical
problems displaying stimulated or spontaneous energy emission,
such as the Vavilov-Cherenkov effect, the anomalous Doppler
effect. When quantum effects were incorporated, it was argued that
rotational superradiance would become a spontaneous process and
that rotating bodies including black holes would slow down by
spontaneous emission of photons. From the historic perspective,
the discovery of black-hole evaporation \cite{HAW} was well
understood from the studies of black-hole superradiance.

Interest in the study of black-hole superradiance has recently
been revived in different areas, including astrophysics,
high-energy physics via the gauge/gravity duality along with
fundamental issues in General Relativity. Superradiant
instabilities can be used to constrain the mass of ultralight
degrees of freedom \cite{INST00, INST0, INST1, INST2}, with
important applications to dark-matter searches. The black hole
superradiance is also associated with the existence of new
asymptotically flat hairy black-hole solutions \cite{HAIR} and
with phase transitions between spinning or charged black objects
and asymptotically anti-de Sitter (AdS) spacetime \cite{ADS0,
ADS1, ADS2} or in higher dimensions \cite{MSHO}. Finally, the
knowledge of superradiance is instrumental in describing the
stability of black holes and in determining the fate of the
gravitational collapse in confining geometries \cite{ADS1}.

During the last few decades, the standard theories of general
relativity have been continuing to explain many important
experimental results. However, there is still some room to the use
of alternative theories of the general theory of relativity. From
a theoretical viewpoint, having an ultraviolet complete theory of
general relativity is complimentary as well as supportive.
Moreover from the observational point of view, general relativity
has shortcomings to describe some gravitational phenomena at a
large scale such as the dark side of the universe. These
shortcomings automatically demand modified theories of general
relativity. The modifications may render some imprints in
astrophysical phenomena where it is expected that the strong
gravity triggers the events in the vicinity of celestial bodies
like astrophysical black holes and neutron stars. Black holes can
be used as a potential probe to investigate the possible
high-energy modifications to general relativity in the regime
where gravity is sufficiently strong. In this respect, the use of
alternative theories of gravity would be of cardinal importance to
study the astrophysical aspects of black hole. The important
astrophysical phenomena namely black hole superradiance is
extremely sensitive to the spacetime geometries linked with it.
Recently, several investigations have been carried out on
superradiance phenomena and on the issues closely linked to it
with the extended framework of modified theories of gravity
\cite{MODG0, MODG1, MODG2, MODG3, MODG4, MODG5, MODG6, MODG7,
MODG8, MODG9, MODG10, MODG11, MODG12, MODG13}. As an extension in
this direction, an attempt has been made here to study the
superradiance of the spinning black holes within the framework of
Lorentz-violating gravity. It is commonly known as the
'Einstein-bumblebee model \cite{KOSTEL0} which involves the
innovative 'spontaneous Lorentz symmetry breaking' principle. From
the theoretical point of view, it arrived from one of the standard
issues of quantization of gravity through string theory. Although
the Lorentz symmetry is the fundamental underlying symmetry of two
successful field theories describing the universe, i.e. GR and the
standard model of particle, however, it is more or less accepted
from all corners that it may break at quantum gravity scales. The
LSB has been introduced through the formulation of an effective
field theory, known as 'standard model extension(SME), where
particle standard model along with GR has been attempted to bring
together in one framework, and every operator is expected to break
the Lorentz symmetry \cite{KOSTEL1, COLL0, COLL1, COLE}. Standard
model extension provides essential inputs to probe LSB both in
high energy particle physics and astrophysics. The SME can be used
in analysis of most modern experimental results indeed. Einstein-
bumblebee model is essentially a simple model that contains
Lorentz symmetry breaking scenario in a significant manner in
which the physical Lorentz symmetry breaks down through an axial
vector field known as the bumblebee field. The breaking of the
Lorentz symmetry in a local Lorentz frame takes place when at
least one quantity carrying local Lorentz indices receives a
non-vanishing vacuum expectation value. In the Einstein bumblebee
model, it is the bumblebee field that receives it. Over the last
few years, a remarkable enthusiasm has been noticed among the
physicist to study the different interesting physical phenomena in
the framework of Einstein Bumblebee model \cite{SEIF, JPAR, DCAP,
CASANA, OLIV, OAV, OAV1, OLIVE, DING}. Recently, the superradiance
phenomenon corresponding to Kerr black hole is studied in
\cite{MK} in this framework. The black hole solution considered
there was Kerr-like. The study of superradiance phenomena of black
holes through the Einstein bumblebee model using Kerr-sen-like
black hole solution is a natural extension. This new investigation
is likely to be useful in the study of black holes in the quantum
gravity realm since it would be possible to compare the
contribution of LV to the superradiance phenomenon.

The article is organized as follows. In Sec.2 a brief discussion
of Einstein-bumblebee gravity with Kerr-Sen-like black hole
solutionis given. A subsection of Sec.2 contains the discussion of
Horizon, Ergosphere, and static limit surface. Sec.3 is devoted
with the superradience scattering of scalar field off
Kerr-Sen-like black hole. Amplification factor  for superradiance
sacttering off Kerr-Sen-like black hole has been calculated in
Sec.4 and a subsection of which is devoted with the superradiant
instability issue for Kerr-Sen-like black hole. In Sec.5
Constraining of the parameter of this black hole is made from the
observed data for $\mathrm{M}87^{*}$. Final Sec. 6 contains a
brief summary and discussions.

\section{EXACT KERR-SEN LIKE BLACK HOLE SOLUTION IN EINSTEIN-BUMBLEBEE MODEL}
Einstein-bumblebee theory is an extension of Einstein's theory
where a vector boson is involved that plays a pivotal role in the
existing symmetry of Einstein's theory \cite{CASANA, ARS, ARS1,
OLIVE, DING, ARS2}. It is an effective classical field theory
where the vector field involved in the theory receives vacuum
expectation when spontaneous braking of an existing symmetry of
the action takes place and a Lorentz violation enters into the
theory as an outcome. Einstein-bumblebee theory is described by
the action
\begin{eqnarray}
\mathcal{S}=\int d^{4} x \sqrt{-g}\left[\frac{1}{16 \pi
G_{N}}\left(\mathcal{R}+\varrho B^{\mu} B^{\nu} \mathcal{R}_{\mu
\nu}\right)-\frac{1}{4} B^{\mu \nu} B_{\mu
\nu}-V\left(B^{\mu}\right)\right]. \label{ACT}
\end{eqnarray}
Here $\varrho^{2}$ stands for the real coupling constant. It
controls the non-minimal gravity interaction to the bumblebee field
$B_{\mu}$ (with the mass dimension 1). The coupling constant
$\varrho^{2}$ has mass dimension $-1$.

The action (\ref{ACT}) leads to the following gravitational field
equation in vacuum
\begin{eqnarray}
\mathcal{R}_{\mu \nu}-\frac{1}{2} g_{\mu \nu} \mathcal{R}=\kappa
T_{\mu \nu}^{B},
\end{eqnarray}
where $\kappa=8 \pi G_{N}$ is the gravitational coupling. The
bumblebee energy momentum tensor $T_{\mu \nu}^{B}$ reads
\begin{eqnarray}\nonumber
T_{\mu \nu}^{B}&=&B_{\mu \alpha} B_{\nu}^{\alpha}-\frac{1}{4}
g_{\mu \nu} B^{\alpha \beta} B_{\alpha \beta}-g_{\mu \nu} V+2
B_{\mu} B_{\nu} V^{\prime} \\\nonumber
&+&\frac{\varrho}{\kappa}[\frac{1}{2} g_{\mu \nu} B^{\alpha}
B^{\beta} R_{\alpha \beta}
-B_{\mu} B^{\alpha} R_{\alpha \nu}-B_{\nu} B^{\alpha} R_{\alpha \mu} \\
&+&\frac{1}{2} \nabla_{\alpha} \nabla_{\mu}\left(B^{\alpha}
B_{\nu}\right)+\frac{1}{2} \nabla_{\alpha}
\nabla_{\nu}\left(B^{\alpha} B_{\mu}\right)-\frac{1}{2}
\nabla^{2}\left(B^{\mu} B_{\nu}\right)-\frac{1}{2} g_{\mu \nu}
\nabla_{\alpha} \nabla_{\beta} \left(B^{\alpha} B^{\beta}\right)].
\end{eqnarray}
Here prime(') denotes the differentiation with respect to the
argument.

Einstein's equation in the present  situation is  generalized to
\begin{eqnarray}\nonumber
&&\mathcal{R}_{\mu \nu}-\kappa b_{\mu \alpha}
b_{\nu}^{\alpha}+\frac{\kappa}{4} g_{\mu \nu} b^{\alpha \beta}
b_{\alpha \beta}+\varrho b_{\mu} b^{\alpha} \mathcal{R}_{\alpha
\nu}+\varrho b_{\nu} b^{\alpha} \mathcal{R}_{\alpha
\mu}\\\nonumber &&-\frac{\varrho}{2} g_{\mu \nu} b^{\alpha}
b^{\beta} \mathcal{R}_{\alpha \beta}
-\frac{\varrho}{2}\left[\nabla_{\alpha}
\nabla_{\mu}\left(b^{\alpha} b_{\nu}\right) +\nabla_{\alpha}
\nabla_{\nu}\left(b^{\alpha} b_{\mu}\right)
-\nabla^{2}\left(b_{\mu} b_{\nu}\right)\right]=0\\
&&\Rightarrow\bar{R}_{\mu \nu}=0, \label{MEQ}
\end{eqnarray}
with
\begin{eqnarray}\nonumber
\bar{R}_{\mu \nu}&=&\mathcal{R}_{\mu \nu}-\kappa b_{\mu \alpha}
b_{\nu}^{\alpha} +\frac{\kappa}{4} g_{\mu \nu} b^{\alpha \beta}
b_{\alpha \beta} +\varrho b_{\mu} b^{\alpha} \mathcal{R}_{\alpha
\nu} +\varrho b_{\nu} b^{\alpha} \mathcal{R}_{\alpha \mu}
-\frac{\varrho}{2} g_{\mu \nu} b^{\alpha} b^{\beta}
\mathcal{R}_{\alpha \beta}
+\bar{B}_{\mu \nu}, \\
\bar{B}_{\mu \nu}&=&-\frac{\varrho}{2}\left[\nabla_{\alpha}
\nabla_{\mu}\left(b^{\alpha} b_{\nu}\right)+\nabla_{\alpha}
\nabla_{\nu}\left(b^{\alpha}
b_{\mu}\right)-\nabla^{2}\left(b_{\mu} b_{\nu}\right)\right].
\end{eqnarray}
If we now adopt the standard Boyer-Lindquist coordinates we
find that the underlying generalized gravity model admits a
Kerr-Sen-like black hole solution:
\begin{eqnarray}
d s^{2}=-\left(1-\frac{2 M r}{\rho^{2}}\right) d t^{2}-\frac{4 M r
a \sqrt{1+\ell} \sin ^{2} \theta}{\rho^{2}} d t d
\varphi+\frac{\rho^{2}}{\Delta} d r^{2}+\rho^{2} d
\theta^{2}+\frac{A \sin ^{2} \theta}{\rho^{2}} d \varphi^{2},
\label{FINAL}
\end{eqnarray}
where
\begin{eqnarray}
 \rho^{2}=r(r+b)+(1+\ell) a^{2} \cos ^{2} \theta,\Delta=\frac{r(r+b)
 -2 M r}{1+\ell}+a^{2}, A=\left[r(r+b)+(1+\ell)
a^{2}\right]^{2}-\Delta(1+\ell)^{2} a^{2} \sin ^{2} \theta.
\end{eqnarray}
Here  $a=\frac{J}{M}$ represents the rotation (Kerr) parameter  $b=\frac{Q^2}{M}$, the Sen parameter related to the electric charge, and  $\ell$ the Lorentz-violating parameter. M, J, and  Q are representing respectively the mass, angular momentum, and charge of the black hole. Note that when  $\ell \rightarrow 0 $  it recovers the usual
Kerr-Sen metric \cite{ASEN}.
\subsection{Horizon, Ergosphere, and static limit surface}
The metric is singular when $\Delta=0$. The roots of the equation
depend on the parameter $M$,$a$, $b$, and $\ell$. Having a maximum
of two real roots, or two equal roots, and no real roots are the
possibilities to occur from the condition$\Delta=0$ \cite{SGG,
SGG1, SGG2, TJDP}. The horizons correspond to the two real roots
which are given by
\begin{eqnarray}
r_{\pm}=M-\frac{b}{2} \pm\frac{
\sqrt{(b-2M)^{2}-4a^{2}(1+\ell)}}{2},
\end{eqnarray}
where $\pm$ signs correspond to the outer and inner horizon
respectively. The event horizon and Cauchy horizon are labelled
with $r_{eh}$ and $r_{ch}$ respectively. We will have a black hole
only when
\begin{equation}
|b-2M|\geq 2a\sqrt{1+l}.
\end{equation}
When $r_{-} = r_{+}$ we have extremal black hole. The parameter
space $(a,b)$ for three different values values of Lorentz
violating parameter $\ell= -0.5, 0$, and $+0.5$ are shown here.

\begin{figure}[H]
\centering
\begin{subfigure}{.52\textwidth}
%\centering
  \includegraphics[width=.8\linewidth]{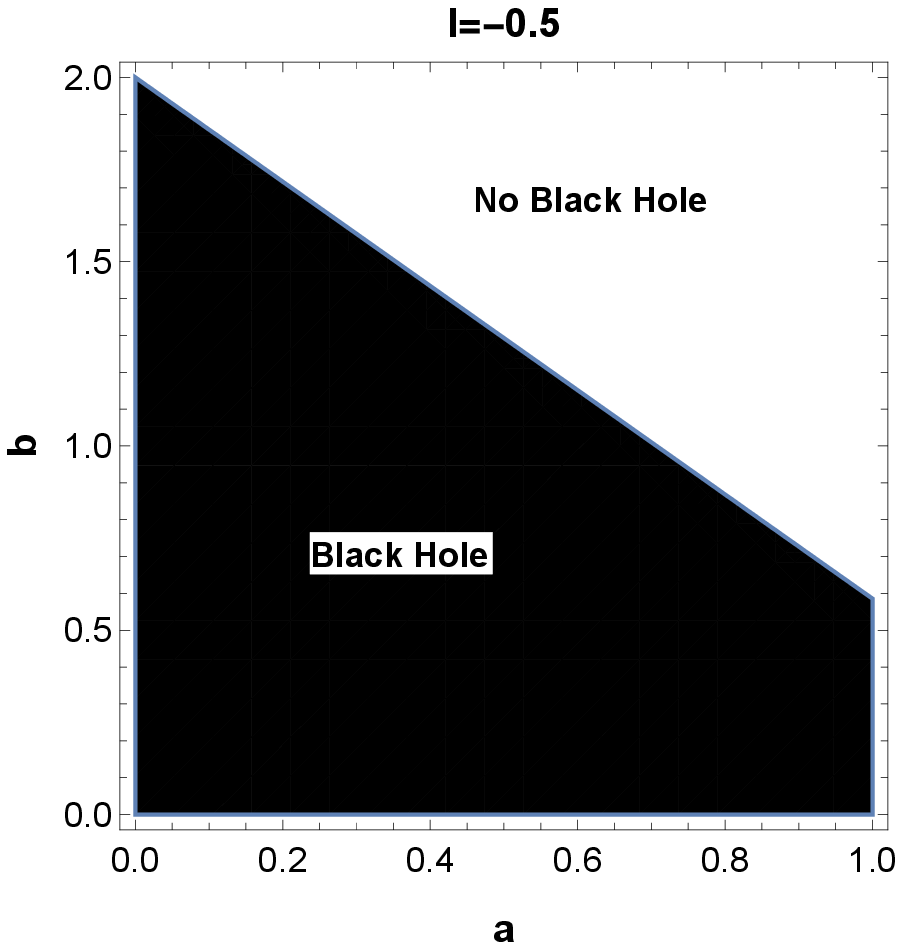}
%\caption{Critical radius for various values of $b$ with $ l=.1,k=.1$ and $\theta=\pi/2$}
\end{subfigure}%
%\hspace{1.5em}%
\begin{subfigure}{.52\textwidth}
\centering
  \includegraphics[width=.8\linewidth]{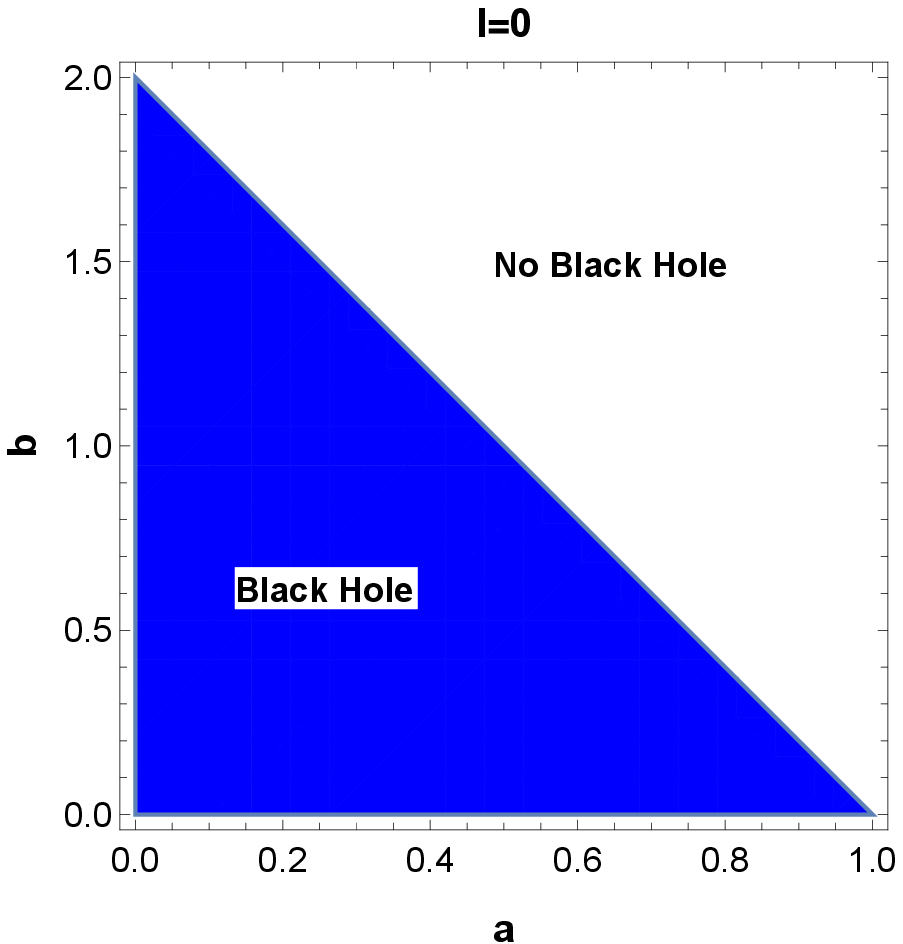}
%\caption{Critical radius for various values of $k$ with $ l=.1,b=.1$ and $\theta=\pi/2$}
\end{subfigure}
%\hspace{1.5em}%
\begin{subfigure}{.52\textwidth}
\centering
  \includegraphics[width=.8\linewidth]{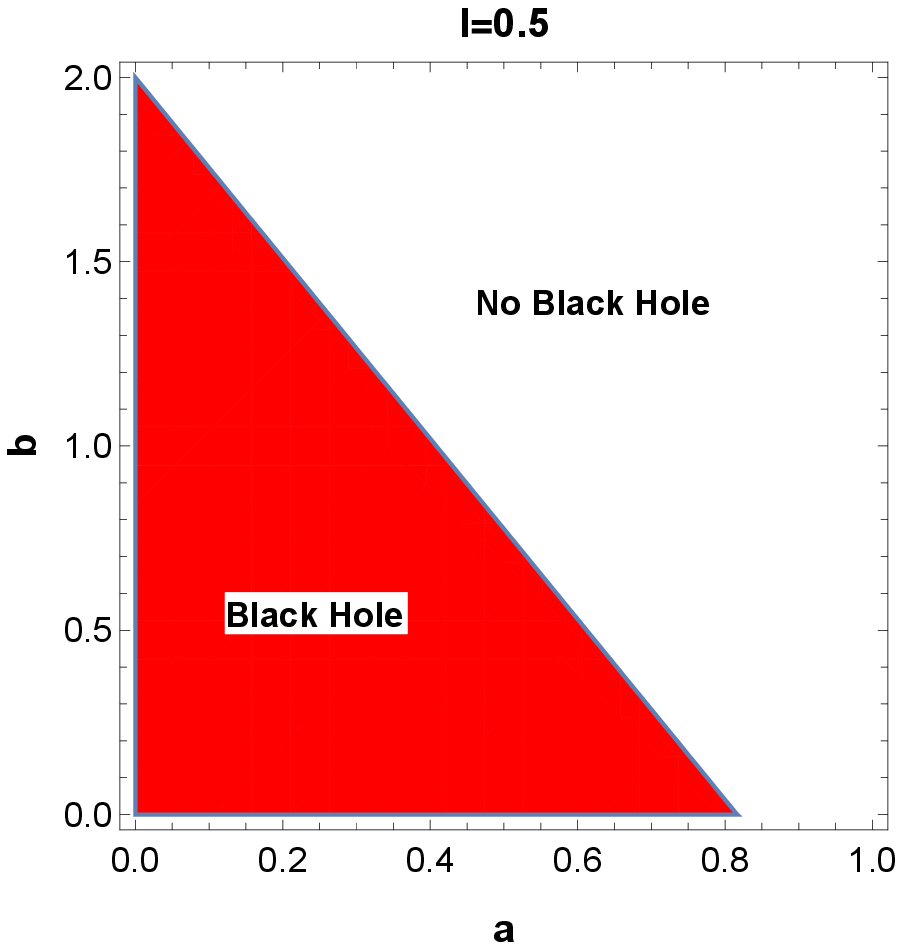}
%\caption{Critical radius for various values of $l$ with $k=.1,b=.1$ and $\theta=\pi/2$}
\end{subfigure}%
\caption{Parameter space($a$,$b$) for various values of $l$.}
\label{fig:test}
\end{figure}

Fig.1 shows that parameter space for which we have a black hole is
shrinking with an increase in the LV parameter $\ell$. So increase
in $\ell$ makes the system less probable for having a black hole
and the reverse is the case when LV parameter $\ell$ decreases. We
now plot the $\Delta$ versus $r/M$ for various different variation
of $a$, $b$, and $\ell$. In the plots when the variation of one
parameter is considered the other two are held fixed.

From Fig. 2 and Fig. 3, we readily observe that there exist a
critical value  $a_{c}$ for the parameter $a$ for fixed values of
$bl$ and $l$. Similarly, there is a critical value $b_{c}$ for the
parameter $b$ when the parameter $a$ and $l$ care kept fixed. For
fixed values of the parameter $b$ and $a$, $l_{c}$ comes out as
critical value for the parameter $\ell$. At these critical values
two roots of the  $\Delta=0$ becomes identical which indicate
extremal black holes. For instance, when $b=0.7, l=0.4$ we have
$a_{c}=0.54935$, when  $a=0.6, l=0.4$ we have $b_{c}=0.580141$,
and when $a=0.6, b=0.7$ we have $l_{c}=0.173611$. Therefore, for
$a < a_{c}$ we have black hole and for $a
> a_{c}$ we have naked singularity. Similarly, $b < b_{c}$
indicates existence of a black hole and for $b > b_{c}$ it is a
naked singularity. For $l < l_{c}$, in a similar way, we have
black hole and $l > l_{c}$ signifies naked singularity.

\begin{figure}[H]
\centering
\begin{subfigure}{.58\textwidth}
%\centering
  \includegraphics[width=.8\linewidth]{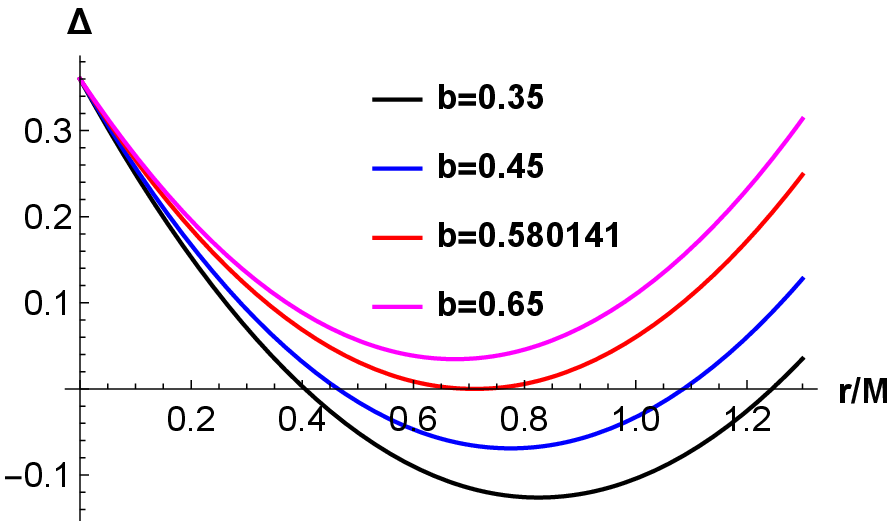}
\end{subfigure}%
\begin{subfigure}{.58\textwidth}
%\centering
  \includegraphics[width=.8\linewidth]{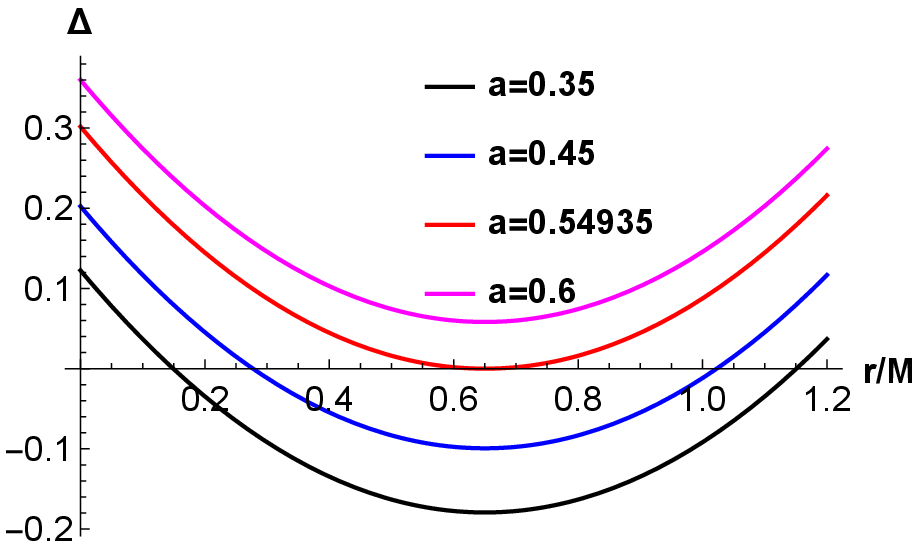}
\end{subfigure}
\caption{The left one gives variation of $\Delta$ for various
values of $b$ with $a=0.6$ and $l=0.4$ and the right one gives
variation for various values of $a$ with $b=0.7$ and $l=0.4$.}
\label{fig:test}
\end{figure}

\begin{figure}[H]
\centering
\begin{subfigure}{.58\textwidth}
  \centering
  \includegraphics[width=.8\linewidth]{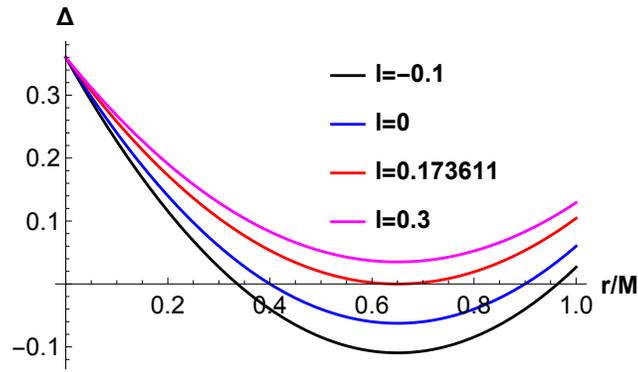}
\end{subfigure}
\caption{It gives variation of $\Delta$for various values of $l$
with $b=0.7$ and $a=0.6$.} \label{fig:test}
\end{figure}

Let us now turn towards the static limit surface (SLS) where the
asymptotic time-translational Killing vector becomes null which
gives
\begin{equation}
g_{tt}=\rho^{2}-2Mr=0.
\end{equation}
The real positive solutions of the above equation give radial
coordinates of the ergosphere given by
\begin{equation}
r_{\pm}^{e r g o}=M
-\frac{b}{2}\pm \frac{\sqrt{(b-2M)^{2}-4a^{2}(1+\ell) \cos ^{2}
\theta}}{2}.
\end{equation}
Inside the SLS no observer can stay static and
they are bound to co-rotate around the black hole. The region
between the SLS and the event
horizon is called the ergosphere shown below in Fig. 4. According to Penrose
 \cite{PENR, PENR1} energy can be extracted from black hole's ergosphere.
\begin{figure}[H]
\centering
\begin{subfigure}{.3\textwidth}
  \centering
  \includegraphics[width=.9\linewidth]{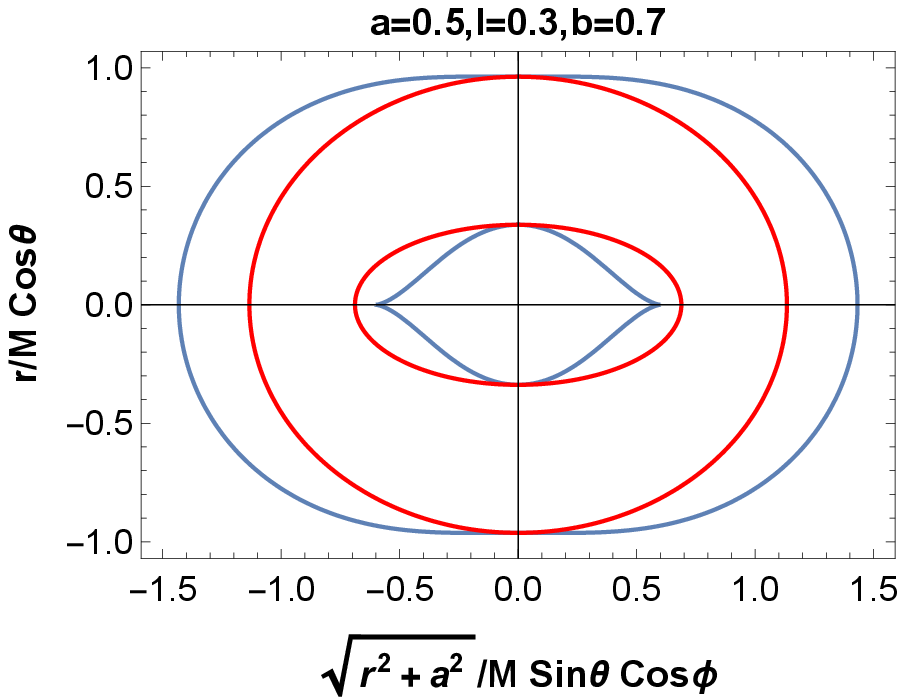}
\end{subfigure}%
\hspace{1.5em}%
\begin{subfigure}{.31\textwidth}
  \centering
  \includegraphics[width=.95\linewidth]{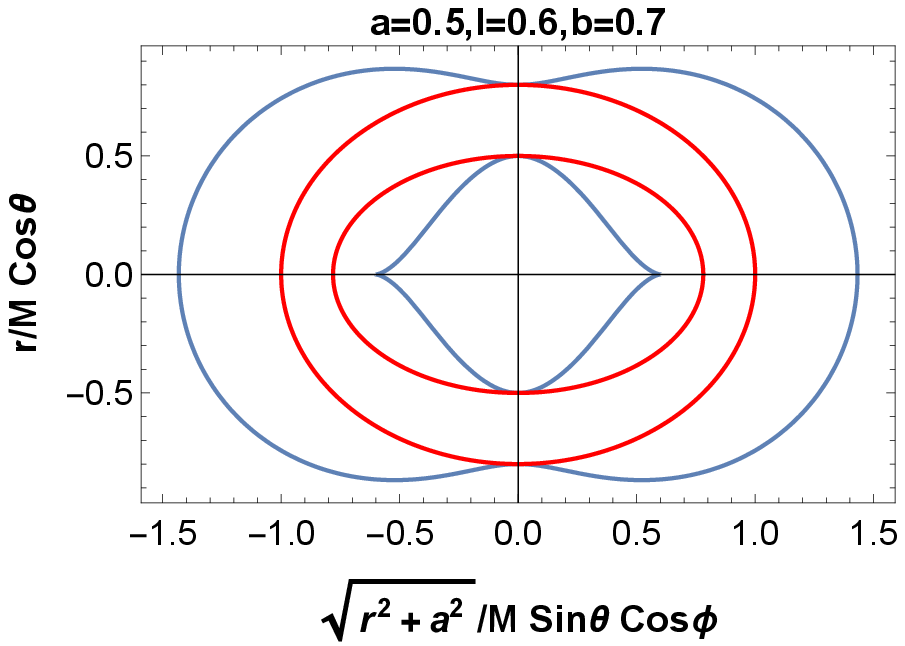}
\end{subfigure}%
\hspace{1.5em}%
\begin{subfigure}{.32\textwidth}
  \centering
  \includegraphics[width=1\linewidth]{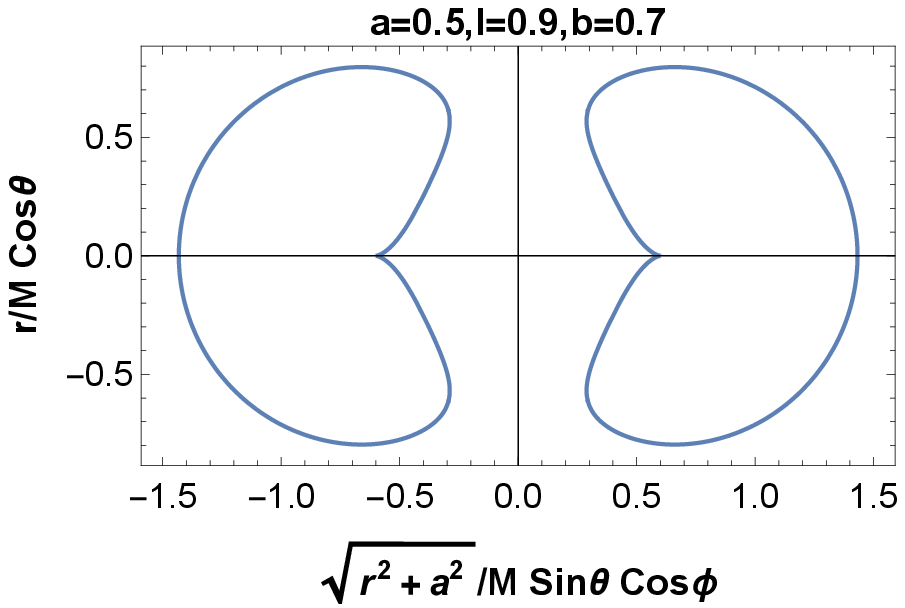}
\end{subfigure}
\par\bigskip
\begin{subfigure}{.3\textwidth}
  \centering
  \includegraphics[width=.9\linewidth]{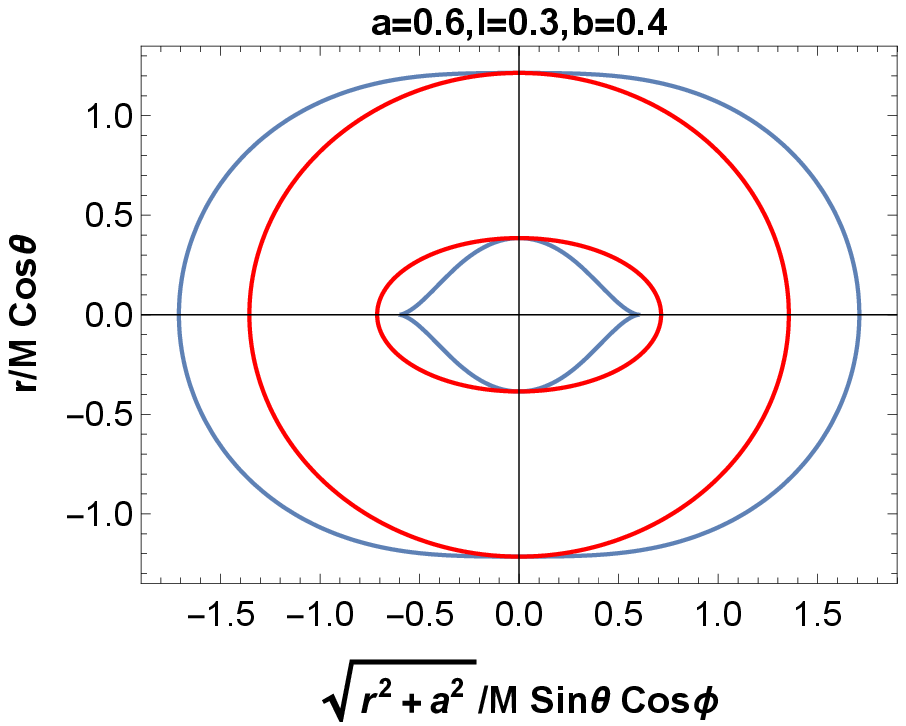}
\end{subfigure}%
\hspace{1.5em}%
\begin{subfigure}{.31\textwidth}
  \centering
  \includegraphics[width=.95\linewidth]{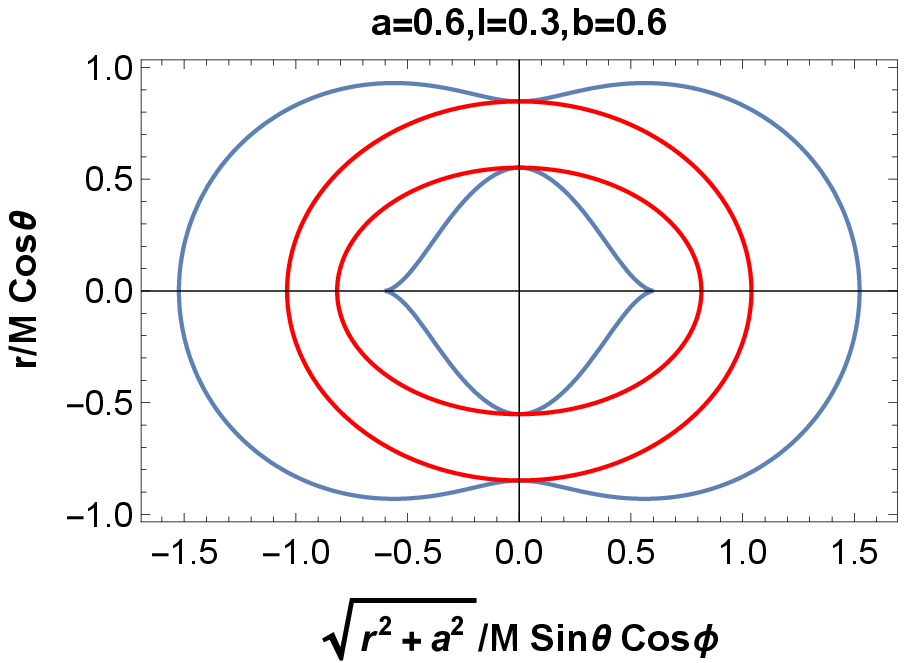}
\end{subfigure}%
\hspace{1.5em}
\begin{subfigure}{.33\textwidth}
  \centering
  \includegraphics[width=1\linewidth]{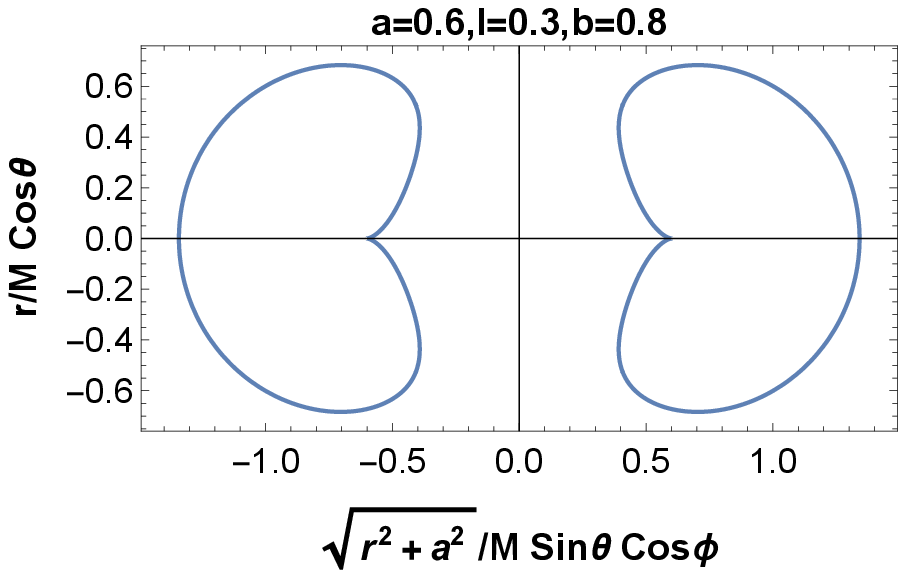}
\end{subfigure}
\caption{The cross-section of event horizon (outer red line),
SLS (outer blue dotted line) and ergoregion of Kerr-Sen like black holes.}
 \label{fig:test}
\end{figure}
Fig. 4 shows that the shape of the ergosphere closely depends on
parameters $a, b,$ and $l$. A careful look reveals that the size
of the ergosphere is enhancing with the increase in the LV
parameter $\ell$ when $a$ and $b$ remain fixed. The value of the
parameter $b$ also has an influence on the size of the ergosphere.
The size of the ergosphere increases with the increase in the
value of the parameter $b$ as well when $a$ and $l$ remain
unchanged.

Horizon angular velocity is found out to be
\begin{equation}
\hat{\Omega}=\frac{a\sqrt{1+l}}{r_{eh}\left(r_{eh}+b\right)+a^{2}\left(1+\ell\right)}.
\end{equation}

\section{Superradience scattering of scalar field off Kerr-Sen-like black hole}
To study the superradiance scattering of a scalar field $\Phi$ with
mass $\mu$ we consider the Klein-Gordon equation in curved
spacetime
\begin{eqnarray}
\left(\bigtriangledown_{\alpha}\bigtriangledown^{\alpha}+\mu^{2}\right)\Phi(t,r,\theta,\phi)
= \left[\frac{-1}{\sqrt{-g}}\partial_{\sigma}\left(g^{\sigma
\tau}\sqrt{-g}\partial_{\tau}\right)+\mu^{2}\right]\Phi(t,r,\theta,\phi)&=&0.
\label{KG}
\end{eqnarray}
Adopting the separation of variables method on the equation
(\ref{KG})  it  is possible to separate it into radial and
angular part using the following ansatz in the standard
Boyer-Lindquist coordinates $(t, r, \theta, \phi)$
\begin{eqnarray}
\Phi(t, r, \theta, \phi)=R_{\omega j m}(r) \Theta(\theta) e^{-i
\omega t} e^{i m \phi}, \quad j \geq 0, \quad-j \leq m \leq j,
\quad \omega>0, \label{PHI}
\end{eqnarray}
where $R_{\omega j m}(r)$ is the radial function and
$\Theta(\theta)$ is the oblate spheroidal wave function. The
symbols $j$, $m$, and $\omega$ respectively stand for the angular
eigenfunction, angular quantum number, and the positive frequency
of the field, which is under investigation, as measured by a far
away observer. Using the the ansatz ({\ref{PHI}) the differential
equation (\ref{KG}), is found to  get separated into the following
two ordinary differential equations. For radial part the equation
reads
\begin{eqnarray}
&&\frac{d}{d r}(\Delta \frac{d R_{\omega j m}(r)}{d
r})+(\frac{((r(r+b)+a^{2}(1+\ell)) \omega-am\sqrt{1+\ell}
)^{2}}{\Delta(1+\ell)})R_{\omega l m}(r)
 \nonumber \\
&&-(\mu^{2} r(r+b)+j(j+1)+a^{2}(1+\ell) \omega^{2}-2 m \omega
a\sqrt{1+\ell}) R_{\omega l m}(r)=0, \label{RE}
\end{eqnarray}
and for the angular part it reads
\begin{eqnarray}
&&\sin \theta \frac{d}{d \theta}\left(\sin \theta \frac{d
\Theta_{\omega j m}(\theta)}{d \theta}\right)+\left(j(j+1) \sin
^{2} \theta-\left(\left(a\sqrt{1+\ell} \omega \sin ^{2}
\theta-m\right)^{2}\right)\right)\Theta_{\omega j m}(\theta)\nonumber \\
&& + a^{2}(1+\ell) \mu^{2} \sin ^{2} \theta \cos ^{2} \theta~
\Theta_{\omega j m}(\theta)=0.
\end{eqnarray}
We can have a general solution of the radial equation (\ref{RE})
using the earlier investigation \cite{BEZERRA, KRANIOTIS}. We have
given it as an appendix-A. However we are intended to study the
scattering of the field $\Phi$  following the articles
\cite{STRO1, STRO2, TEUK, PAGE}, and in this situation we have
used the asymptotic matching procedure which is explicitely used
in \cite{RAN}. This article however is an extension of the
important works \cite{STRO1, STRO2, TEUK, PAGE}. Use of asymptotic
matching also has been found in \cite{MK}. This led us to land
onto the required result without using the general solution. Let
us first focus on the radial equation. To deal with the radial
equation according to our need we apply a Regge-Wheeler-like
coordinate $r_{*}$ which is defined by
\begin{eqnarray}
r_{*} \equiv \int d r \frac{r(r+b)+a^{2}(1+\ell)}{\Delta},
\quad\left(r_{*} \rightarrow-\infty \quad \text{at event horizon},
\quad r_{*} \rightarrow \infty \quad \text{at infinity} \right)
\end{eqnarray}
To transform the equation into the desired shape, we introduce a
new radial function $\mathcal{R}_{\omega j
m}\left(r_{*}\right)=\sqrt{r(r+b)+a^{2}(1+\ell)} R_{\omega j
m}(r)$. After a few steps of algebra, we obtain the radial
equation with our desired form where an effective potential took
its entry into the picture.
\begin{equation}
\frac{d^{2} \mathcal{R}_{\omega l m}\left(r_{*}\right)}{d
r_{*}^{2}}+V_{\omega j m}(r) \mathcal{R}_{\omega j
m}\left(r_{*}\right)=0. \label{RE1}
\end{equation}
The effective potential that has the crucial role on the
scattering reads
\begin{eqnarray}
%\begin{split}
V_{\omega j m}(r)&=&\frac{1}{1+\ell}\left(\omega-\frac{m
\hat{a}}{r(r+b)+\hat{a}^{2}}\right)^{2}-\frac{\Delta}{\left(r(r+b)
+\hat{a}^{2}\right)^{2}}\left[\right. j(j+1)+\hat{a}^{2}
\omega^{2}-2 m \hat{a} \omega+\mu^{2} r(r+b)
\\\nonumber
&&\left.+\sqrt{r(r+b)+\hat{a}^{2}}\frac{d}{dr}\left(\frac{\Delta
(2r+b)}{2\left(r(r+b)+\hat{a}^{2}\right)^{\frac{3}{2}}}\right)\right],
\label{POT}
%\end{split}
\end{eqnarray}
where $\tilde{a}=a(1+\ell)^{\frac{1}{2}}$. We are intended to
study the scattering of the scalar field $\Phi$ under this
effective potential. In this context, it is beneficial to study
the asymptotic behavior of the scattering potential at the event
horizon and at spatial infinity. In the  asymptotic limit the
potential at the event horizon looks
\begin{eqnarray}
\lim _{r \rightarrow r_{eh}} V_{\omega j
m}(r)=\frac{1}{1+\ell}\left(\omega-m \tilde{\Omega}_{h}\right)^{2}
\equiv k_{e h}^{2},
\end{eqnarray}
%\end{document} and the same at spatial infinity reads
\begin{equation}
\lim _{r \rightarrow \infty} V_{\omega j m}(r)=\omega^{2}-\lim _{r
\rightarrow \infty} \frac{\mu^{2} r(r+b)
\Delta}{\left(r(r+b)+\tilde{a}^{2}\right)^{2}}=\frac{\omega^{2}}{1+\ell}
-\hat{\mu}^{2}\equiv k_{\infty}^{2}, ~~~
\hat{\mu}=\frac{\mu}{\sqrt{\ell+1}}.
\end{equation}
Note that at the two extremal points, event horizon and spatial
infinity, the potential asymptotically shows constant behavior.
However, the values of the constants are different indeed.

We are now in a position to see  the asymptotic  behavior of  the
radial equation.  It is found that the radial equation (\ref{RE1})
has the following asymptotic solutions
\begin{equation}\label{AS}
R_{\omega j m}(r) \rightarrow\left\{\begin{array}{cl}
\frac{A_{i n}^{eh} e^{-i k_{eh} r_{*}}}{\sqrt{r_{e h}(r_{eh}+b)
+\hat{a}^{2}}} & \text { for } r \rightarrow r_{e h} \\
\mathcal{A}_{i n}^{\infty} \frac{e^{-i k_{\infty}
r_{*}}}{r}+\mathcal{A}_{r e f}^{\infty} \frac{e^{i k_{\infty}
r_{*}}}{r} & \text { for } r \rightarrow \infty
\end{array}\right.
\end{equation}
Here $\mathcal{A}_{in}^{eh}$ represents the amplitude of the
incoming scalar wave at event horizon("eh"), and
$\mathcal{A}_{in}^{\infty}$ is the corresponding quantity of the
incoming scalar wave at infinity$("\infty")$. Along with these,
the amplitude of the reflected part of  scalar wave at infinity
$("\infty")$ is $\mathcal{A}_{ref}^{\infty}$.

Let us now compute the Wronskian for the region adjacent to the
event horizon and at infinity. It is found that  Wronskian for
this region is
\begin{equation}
W_{eh}=\left(R_{\omega j m}^{e h} \frac{d R_{\omega j m}^{* e
h}}{d r_{*}}-R_{\omega j m}^{* eh} \frac{d R_{\omega j m}^{eh}}{d
r_{*}}\right),
\end{equation}
and the Wronskian at infinity  reads
\begin{equation}
W_{\infty}=\left(R_{\omega j m}^{\infty} \frac{d R_{\omega j m}^{*
\infty}}{d r_{*}}-R_{\omega j m}^{* \infty} \frac{d R_{\omega j
m}^{\infty}}{d r_{*}}\right).
\end{equation}
The solutions  are linearly independent. From the knowledge of
standard theory of ordinary differential equation it can be
understandable that the Wronskian corresponding to the solutions
will be independent of $r^*$. Thus, the Wronskian evaluated at
horizon does amenable  to equate with the Wronskian evaluated at
infinity. In physical sense, it is reflecting the flux
conservation \cite{REVIEW}. It results an important relation
between the amplitudes of incoming and reflected waves at
different regions of interest.
\begin{equation}
\left|\mathcal{A}_{r e f}^{\infty}\right|^{2}=\left|\mathcal{A}_{i
n}^{\infty}\right|^{2}-\frac{k_{e
h}}{k_{\infty}}\left|\mathcal{A}_{i n}^{e h}\right|^{2}.
\label{AMP}
\end{equation}
The above equation transpires that if $\frac{k_{e
h}}{k_{\infty}}<0$ i.e.,  $\omega<m \hat{\Omega}_{e h}$, the
scalar wave will be superradiantly amplified, because in this
situation, the relation  $\left|\mathcal{A}_{r e
f}^{\infty}\right|^{2}>\left|\mathcal{A}_{i
n}^{\infty}\right|^{2}$  holds explicitly.
\section{Amplification factor $Z_{jm}$ for superradiance}
We now rewrite the radial equation (\ref{RE}) as
\begin{eqnarray}\nonumber
&&\Delta^{2} \frac{d^{2} R_{\omega j m}(r)}{d r^{2}}+\Delta
\frac{d \Delta}{d r} \cdot \frac{d R_{\omega j m}(r)}{d r}\\
 &&+\left(\frac{\left(\left(r(r+b)+\hat{a}^{2}\right)
\omega-\hat{a} m\right)^{2}}{1+\ell}-\Delta\left(\mu^{2}
r(r+b)+j(j+1)+\hat{a}^{2} \omega^{2}-2 m \hat{a}
\omega\right)\right) R_{\omega j m}(r)=0. \label{RE2}
\end{eqnarray}
We now turn to  derive the near-region as well as  the far-region
solution and try to find out a single solution matching the
near-region solution at infinitely with the far-region solution at
its initial point such that this single solution works in the
vicinity of the cardinal region. We apply the change of variable
$x=\frac{r-r_{eh}}{r_{eh}-r_{ch}}$. Using this change of variable
equation (\ref{RE2}) under the approximation $\hat{a} \omega \ll
1$ turns into
\begin{eqnarray}
&&\frac{x^{2}(x+1)^{2}}{(\ell+1)^{2}} \frac{\mathrm{d}^{2}
R_{\omega j m}(x)}{\mathrm{d} x^{2}}+\frac{x(x+1)(2
x+1)}{(\ell+1)^{2}} \frac{\mathrm{d} R_{\omega j m}(x)}{\mathrm{d}
x} \\\nonumber &&+\left(\frac{P^2
x^{4}}{1+\ell}+\frac{B^{2}}{1+\ell}-\frac{j(j+1)}{\ell+1}
x(x+1)-\frac{\hat{\mu}^{2} P^{2}}{\omega^{2}}
x^{3}(x+1)-\hat{\mu}^{2} r_{e h}^{2} x(x+1)-\frac{2 \hat{\mu}^{2}
r_{e h} P}{\omega} x^{2}(x+1) \right. \\\nonumber
&&\left.-\frac{\hat{\mu}^{2}Pb}{\omega}x^{2}(1+x)-\hat{\mu}^{2}br_{eh}x(1+x)\right)
R_{\omega j m}(x)=0,
\end{eqnarray}
where $P=\left(r_{e h}-r_{c h}\right) \omega$ and
$B=\frac{(\omega-m \hat{\Omega})}{r_{e h}-r_{c h}} r_{e h}^{2}$.
For near-region we have  $P x \ll 1$ and $\hat{\mu}^{2} r_{e
h}^{2} \ll 1$ and hence the above equation reduces to
\begin{eqnarray}
x^{2}(x+1)^{2} \frac{\mathrm{d}^{2} R_{\omega j m}(r)}{\mathrm{d}
x^{2}}+x(x+1)(2 x+1) \frac{\mathrm{d} R_{\omega j
m}(r)}{\mathrm{d} x}+\left((\ell+1)B^{2}-j(j+1)(\ell+1)
x(x+1)\right) R_{\omega j m}(r)=0.
\end{eqnarray}
The approximation $\left(\hat{\mu}^{2} r_{e h}^{2} \ll 1\right)$
is originated from the consideration that the Compton wavelength
of the boson participating in the scattering process is much
smaller than the size of the black hole. The general solution of
the above equation in terms of associated Legendre function of the first kind
$P_{\lambda}^{\nu}(y)$ can be written down as
\begin{eqnarray}
R_{\omega j m}(x)=c
P^{2i\sqrt{1+\ell}B}_{\frac{\sqrt{1+4j(j+1)
(l+1)}-1}{2}}(1+2x).
\end{eqnarray}
We now use the relation
\begin{equation}
P_{\lambda}^{\nu}(z)=\frac{1}{\Gamma(1-\nu)}\left(\frac{1+z}{1-z}\right)^{\nu
/ 2}{ }_{2} F_{1}\left(-\lambda, \lambda+1 ; 1-\nu ;
\frac{1-z}{2}\right).
\end{equation}
It enables us to  express $R_{\omega j m}(x)$ in terms of the
ordinary hypergeometric functions ${ }_{2} F_{1}(a, b ; c ; z)$ :
\begin{equation}
R_{\omega j m}(x)=c\left(\frac{x}{x+1}\right)^{-i\sqrt{\ell+1} B}{
}_{2} F_{1}\left(\frac{1-\sqrt{1+4(\ell+1) j(j+1)}}{2},
\frac{1+\sqrt{1+4(\ell+1) j(j+1)}}{2} ; 1-2 i\sqrt{\ell+1} B
;-x\right).\label{NEAR}
\end{equation}
As we have mentioned, we require a single solution using the
matching condition at the desired position where the two solutions
mingle with each other. In this respect, we need to observe the
large $x$ behavior of the above expression. The Eqn. (\ref{NEAR})
for large x $(x \to\infty$) turns into
\begin{eqnarray}
R_{\text {near-large } x} \sim c &&\left(\frac{\Gamma(\sqrt{1+4(\ell+1) j(j+1)})
 \Gamma(1-2 i\sqrt{\ell+1} B)}{\Gamma\left(\frac{1+\sqrt{1+4(\ell+1) j(j+1)}}{2}
 -2 i\sqrt{\ell+1} B\right) \Gamma\left(\frac{1+\sqrt{1+4(\ell+1) j(j+1)}}{2}\right)}
 x^{\frac{\sqrt{1+4(\ell+1) j(j+1)}-1}{2}}+\right.\\
&& \frac{\Gamma(-\sqrt{1+4(\ell+1) j(j+1)}) \Gamma(1-2
i\sqrt{\ell+1} B)}{\Gamma\left(\frac{1-\sqrt{1+4(\ell+1)
j(j+1)}}{2}\right) \Gamma\left(\frac{1-\sqrt{1+4(\ell+1)
j(j+1)}}{2}-2 i\sqrt{\ell+1} B\right)}
x^{\left.-\frac{\sqrt{1+4(\ell+1) j(j+1)}+1}{2}\right)} .
\label{NF}
\end{eqnarray}
For the far-region, we can use the  approximations $x+1 \approx x$
and $\hat{\mu}^{2} r_{e h}^{2} \ll 1$. We may drop all the terms
except those which describe the free motion with momentum $j$ and
that reduces equation (\ref{RE2}) to
\begin{equation}
\frac{\mathrm{d}^{2} R_{\omega j m}(x)}{\mathrm{d}
x^{2}}+\frac{2}{x} \frac{\mathrm{d} R_{\omega j m}(x)}{\mathrm{d}
x}+\left(k_{l}^{2}-\frac{j(j+1)(\ell+1)}{x^{2}}\right) R_{\omega j
m}(x)=0, \label{FAR}
\end{equation}
where $k_{l} \equiv \frac{P\sqrt{1+\ell}}{\omega}
\sqrt{\omega^{2}-\mu^{2}}$. Equation (\ref{FAR}) has the
general solution
\begin{eqnarray}\nonumber
R_{\omega j m, \text { far }}=e^{-i k x}\left(d_{1} x^{\frac{\sqrt{1+4(\ell+1)
 j(j+1)}-1}{2}} M\left(\frac{1+\sqrt{1+4(\ell+1) j(j+1)}}{2},
 1+\sqrt{1+4(\ell+1) l(l+1)}, 2 i k_{l} x\right)+\right. \\
\left.d_{2} x^{-\frac{\sqrt{1+4(\alpha+1) j(j+1)}+1}{2}}
M\left(\frac{1-\sqrt{1+4(\ell+1) j(j+1)}}{2}, 1-\sqrt{1+4(\ell+1)
j(j+1)}, 2 i k_{l} x\right)\right), \label{FARR}
\end{eqnarray}
where $M(a, b, y)$ refers to the confluent hypergeometric Kummer
function of first kind. In order to match the solution with
(\ref{NF}), we look for  the  small $x$ behavior of the solution
(\ref{FARR}). For small $x (x \to 0)$, the equation (\ref{FARR})
takes the form
\begin{equation} R_{\omega
j m, \text { far-small } \mathrm{x}} \sim d_{1}
x^{\frac{\sqrt{1+4(\ell+1) j(j+1)}-1}{2}}+d_{2}
x^{-\frac{1+\sqrt{1+4(\ell+1) j(j+1)}}{2}}. \label{FN}
\end{equation}
The solution (\ref{NF}) and (\ref{FN}) are susceptible for
matching, since these two have common region of interest. The
matching of the asymptotic solutions (\ref{NF}) and (\ref{FN})
enables us to compute the scalar wave flux at infinity resulting in
\begin{eqnarray}
d_{1}=&\left.c \frac{\Gamma(\sqrt{1+4(\ell+1) j(j+1)})
\Gamma(1-2 i\sqrt{\ell+1} B)}{\left.\Gamma\left(\frac{1+\sqrt{1+4(\ell+1) j(j+1)}}{2}\right)
-2 i\sqrt{\alpha+1} B\right) \Gamma\left(\frac{1+\sqrt{1+4(\ell+1) j(j+1)}}{2}\right.}\right), \\\
d_{2}=& c \frac{\Gamma(-\sqrt{1+4(\ell+1) j(j+1)}) \Gamma(1-2
i\sqrt{\ell+1} B)}{\left.\Gamma\left(\frac{1-\sqrt{1+4(\ell+1)
j(j+1)}}{2}\right)-2 i\sqrt{\alpha+1} B\right)
\Gamma\left(\frac{1-\sqrt{1+4(\ell+1) j(j+1)}}{2}\right)}.
\label{DD}
\end{eqnarray}
We expand equation (\ref{FARR}) around infinity which after
expansion results
\begin{eqnarray}
d_{1} \frac{\Gamma(1+\sqrt{1+4(\ell+1) j(j+1)})}{\Gamma\left(\frac{1+\sqrt{1+4(\ell+1)
j(j+1)}}{2}\right)} k_{l}^{-\frac{1+\sqrt{1+4(\ell+1)
j(j+1)}}{2}}\left((-2 i)^{-\frac{1+\sqrt{1+4(\ell+1)
 j(j+1)}}{2}} \frac{e^{-i k_{l} x}}{x}+(2 i)^{-\frac{1+\sqrt{1+4(\ell+1) j(j+1)}}{2}}
  \frac{e^{i k_{l} x}}{x}\right)+ \\\nonumber
d_{2} \frac{\Gamma(1-\sqrt{1+4(\ell+1)
j(j+1)})}{\frac{1-\sqrt{1+4(\ell+1) j(j+1)}}{2}}
k_{l}^{\frac{\sqrt{1+4(\ell+1) j(j+1)}-1}{2}}\left((-2
i)^{\frac{\sqrt{1+4(\ell+1) j(j+1)}-1}{2}} \frac{e^{-i k_{l}
x}}{x}+(2 i)^{\frac{\sqrt{1+4(\ell+1) j(j+1)}-1}{2}} \frac{e^{i
k_{l} x}}{x}\right).
\end{eqnarray}
With the approximations $\frac{1}{x} \sim \frac{P}{\omega} \cdot
\frac{1}{r}, \quad e^{\pm i k_{l} x} \sim e^{\pm i
\sqrt{(1+\ell)(\omega^{2}-\mu^{2})} r}$, if we match the above
solution with the radial solution \eqref{AS}
$$
R_{\infty}(r) \sim \mathcal{A}_{i n}^{\infty} \frac{e^{-i
\sqrt{\frac{\omega^{2}}{1+\ell}-\hat{\mu}^{2}}
r^{*}}}{r}+\mathcal{A}_{r e f}^{\infty} \frac{e^{i
\sqrt{\frac{\omega^{2}}{1+\ell}-\hat{\mu}^{2}} r^{*}}}{r}, \quad
\text { for } \quad r \rightarrow \infty
$$
we get
$$
\begin{array}{c}
\mathcal{A}_{i n}^{\infty}=\frac{P}{\omega}\left(d_{1}(-2 i)^{-\frac{1+\sqrt{1+4(\ell+1)j(j+1)}}{2}}
 \frac{\Gamma(1+\sqrt{1+4(\ell+1)j(j+1)})}{\Gamma\left(\frac{1+\sqrt{1+4(\ell+1)j(j+1)}}{2}\right)}
 k_{l}^{-\frac{1+\sqrt{1+4(\ell+1)j(j+1)}}{2}}+\right. \\
\left.d_{2}(-2 i)^{\frac{\sqrt{1+4(\ell+1)j(j+1)}-1}{2}}
\frac{\Gamma(1-\sqrt{1+4(\ell+1)j(j+1)})}{\Gamma\left(\frac{1-\sqrt{1+4(\ell+1)j(j+1)}}{2}\right)}
k_{l}^{\frac{\sqrt{1+4(\ell+1)j(j+1)}-1}{2}}\right),
\end{array}
$$
and
$$
\begin{array}{l}
\mathcal{A}_{r e f}^{\infty}=\frac{P}{\omega}\left(d_{1}(2 i)^{-\frac{1+\sqrt{1+4(\ell+1)j(j+1)}}{2}}
 \frac{\Gamma(1+\sqrt{1+4(\ell+1)j(j+1)})}{\Gamma\left(\frac{1+\sqrt{1+4(\ell+1)j(j+1)}}{2}\right)}
  k_{l}^{-\frac{1+\sqrt{1+4(\ell+1)j(j+1)}}{2}}+\right. \\
\left.d_{2}(2 i) \frac{\sqrt{1+4(\ell+1)j(j+1)}-1}{2}
\frac{\Gamma(1-\sqrt{1+4(\ell+1)j(j+1)})}{\Gamma\left(\frac{1-\sqrt{1+4(\ell+1)j(j+1)}}{2}\right)}
k_{l}^{\frac{\sqrt{1+4(\ell+1)j(j+1)}-1}{2}}\right) .
\end{array}
$$
Substituting  the expressions of $d_{1}$ and $d_{2}$ from Eqn.
(\ref{DD}) into the above expressions we have
\begin{eqnarray}
\mathcal{A}_{in}^{\infty}&=&\frac{c(-2
i)^{-\frac{1+\sqrt{1+4(\ell+1)
j(j+1)}}{2}}}{\sqrt{(1+\ell)(\omega^{2}-\mu^{2})}} \cdot
\frac{\Gamma(\sqrt{1+4(\ell+1) j(j+1)}) \Gamma(1+\sqrt{1+4(\ell+1)
j(j+1)})}{\Gamma\left(\frac{1+\sqrt{1+4(\ell+1)j(j+1)}}{2}-2
i\sqrt{\ell+1}
B\right)\left(\Gamma\left(\frac{1+\sqrt{1+4(\ell+1)j(j+1)}}{2}\right)\right)^{2}}\times
\\\nonumber &&\Gamma(1-2 i\sqrt{\alpha+1} B)
k_{l}^{\frac{1-\sqrt{1+4(\ell+1)j(j+1)}}{2}}+\frac{c(-2
i)^{\frac{\sqrt{1+4(\ell+1)j(j+1)}-1}{2}}}{\sqrt{(1+\ell)(\omega^{2}-\hat{\mu}^{2})}}
\times \\\nonumber &&\frac{\Gamma(1-\sqrt{1+4(\ell+1)j(j+1)})
\Gamma(-\sqrt{1+4(\ell+1)j(j+1)})}{\left(\Gamma\left(\frac{1-\sqrt{1+4(\ell+1)j(j+1)}}{2}\right)\right)^{2}
\Gamma\left(\frac{1-\sqrt{1+4(\ell+1)j(j+1)}}{2}-2 i\sqrt{\ell+1}
B\right)} \Gamma(1-2 i\sqrt{\ell+1} B)
k_{l}^{\frac{1+\sqrt{1+4(\ell+1)j(j+1)}}{2}},
\end{eqnarray}
and
\begin{eqnarray}
\mathcal{A}_{ref}^{\infty}&=&\frac{c(2
i)^{-\frac{1+\sqrt{1+4(\ell+1)
j(j+1)}}{2}}}{\sqrt{(1+\ell)(\omega^{2}-\mu^{2})}} \cdot
\frac{\Gamma(\sqrt{1+4(\ell+1) j(j+1)}) \Gamma(1+\sqrt{1+4(\ell+1)
j(j+1)})}{\Gamma\left(\frac{1+\sqrt{1+4(\ell+1)j(j+1)}}{2}-2
i\sqrt{\ell+1}
B\right)\left(\Gamma\left(\frac{1+\sqrt{1+4(\ell+1)j(j+1)}}{2}\right)\right)^{2}}\times
\\\nonumber &&\Gamma(1-2 i\sqrt{\alpha+1} B)
k_{l}^{\frac{1-\sqrt{1+4(\ell+1)j(j+1)}}{2}}+\frac{c(2
i)^{\frac{\sqrt{1+4(\ell+1)j(j+1)}-1}{2}}}{\sqrt{(1+\ell)(\omega^{2}-\hat{\mu}^{2})}}
\times \\\nonumber &&\frac{\Gamma(1-\sqrt{1+4(\ell+1)j(j+1)})
\Gamma(-\sqrt{1+4(\ell+1)j(j+1)})}{\left(\Gamma\left(\frac{1-\sqrt{1+4(\ell+1)j(j+1)}}{2}\right)\right)^{2}
\Gamma\left(\frac{1-\sqrt{1+4(\ell+1)j(j+1)}}{2}-2 i\sqrt{\ell+1}
B\right)} \Gamma(1-2 i\sqrt{\ell+1} B)
k_{l}^{\frac{1+\sqrt{1+4(\ell+1)j(j+1)}}{2}}.
\end{eqnarray}
The amplification factor ultimately results out to be
\begin{equation}
Z_{j m} \equiv \frac{\left|\mathcal{A}_{r e
f}^{\infty}\right|^{2}}{\left|\mathcal{A}_{i
n}^{\infty}\right|^{2}}-1. \label{AMPZ}
\end{equation}
Equation (\ref{AMPZ}) is a general expression of the amplification
factor obtained by making use of the asymptotic matching method.
When
$\frac{\left|\mathcal{A}_{ref}^{\infty}\right|^{2}}{\left|\mathcal{A}_{i
n}^{\infty}\right|^{2}}$ acquires a value greater than unity there
will be a gain in amplification factor that corresponds to
superradiance phenomena. However, a negative value of the
amplification factor indicates a loss that corresponds to the
nonappearance of superradiance. To study the effect of Lorentz
violation on the superradiance phenomena, it will be useful to
plot $Z_{j m}$ versus $M\omega$ for different LV parameters. In
Fig. (6), we present the variation $Z_{j m}$ versus $M\omega$ for the
leading multipoles $J= 1$, and $2$ taking different values (both
negative and positive) of LV Parameter. From the fig. (5) along with fig. (6), it is
evident that superradiance for a particular $j$ occurs when the
allowed values of $m$ are restricted to $m > 0$.

For negative $m$ amplification factor takes negative value which refers to the
nonoccurrence of superradiance. The plots also show transparently
that with the decrease in the value of the LV paraneters the
superradiance process enhances and the reverse is the case when
the value of the LV parameter decreases. In Fig. (8) we have also
studied the effect of the parameter $b=\frac{Q^2}{M}$ on the
superradiance scenario. It shows that the superradiance scenario
gets diminished with the increase in the value of the parameter
$b$. In \cite{ARS} we have noticed that the size of the shadow
decreases with the increase in the value of both the parameters
$l$ and $b$. The only difference is that $l$ can take both
positive values, however, $b$ as per definition can not be
negative. Therefore, an indirect relation of superradiance with
the size of the shadow is being revealed through this analysis. A
decrease in the value of $b$ and $l$ indicate the increase in the
size of the shadow.

\begin{figure}[H]
\centering
\begin{subfigure}{.58\textwidth}
%\centering
  \includegraphics[width=.8\linewidth]{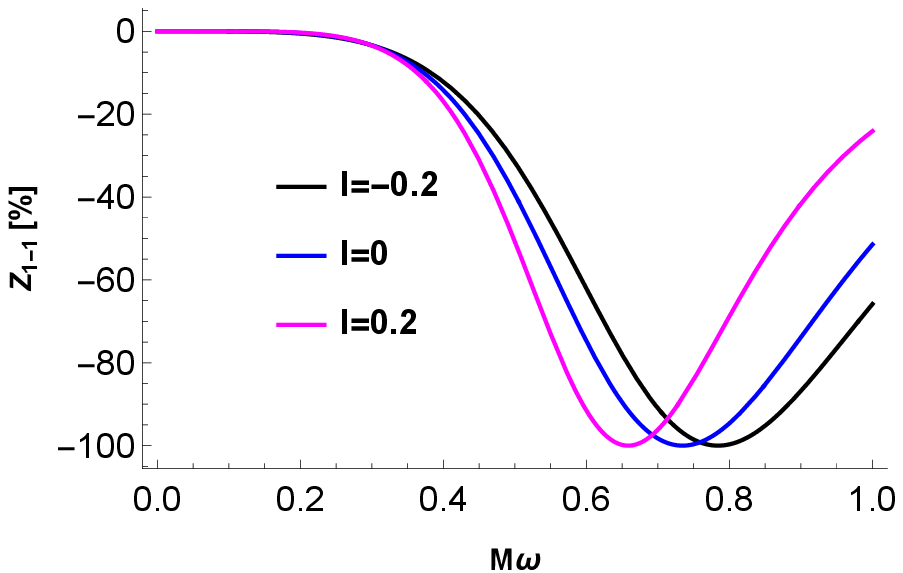}
%\caption{Critical radius for various values of $b$ with $ l=.1,k=.1$ and $\theta=\pi/2$}\hspace{1em}%
\end{subfigure}%
\begin{subfigure}{.58\textwidth}
%\centering
  \includegraphics[width=.8\linewidth]{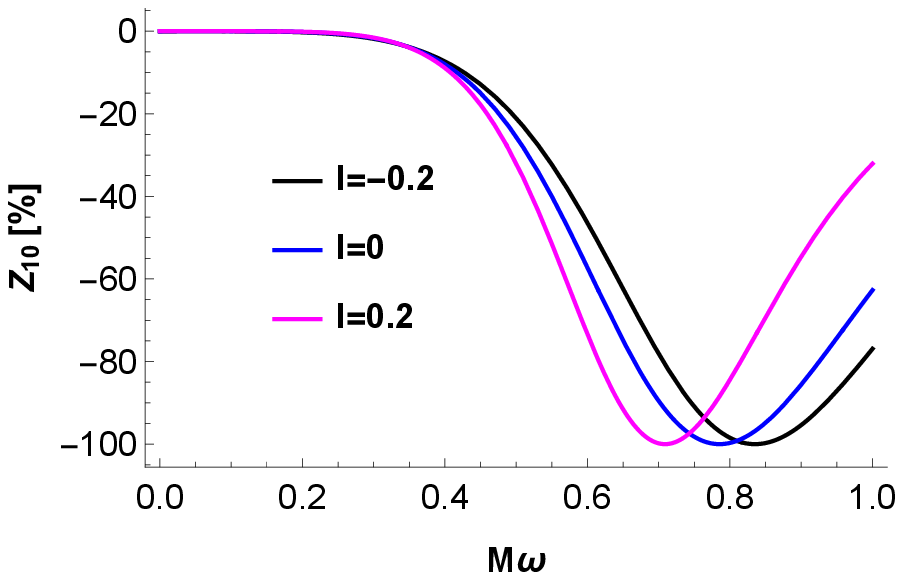}
%\caption{Critical radius for various values of $k$ with $ l=.1,b=.1$ and $\theta=\pi/2$}\hfill
\end{subfigure}
\caption{Variation of amplification factors with $\ell$ for
non-superradiant multipoles with $\hat{\mu}=0.1,b=0.1$, and
$\hat{a}=0.4$.} \label{fig:test}
\end{figure}

\begin{figure}[H]
\centering
\begin{subfigure}{.58\textwidth}
%\centering
  \includegraphics[width=.8\linewidth]{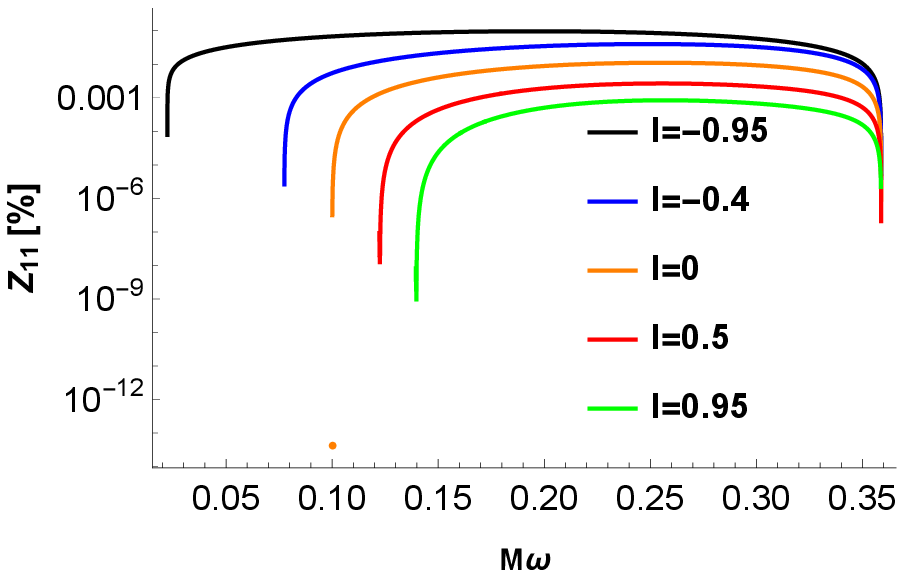}
%\caption{Critical radius for various values of $b$ with $ l=.1,k=.1$ and $\theta=\pi/2$}
\end{subfigure}%
\begin{subfigure}{.58\textwidth}
 %\centering
  \includegraphics[width=.8\linewidth]{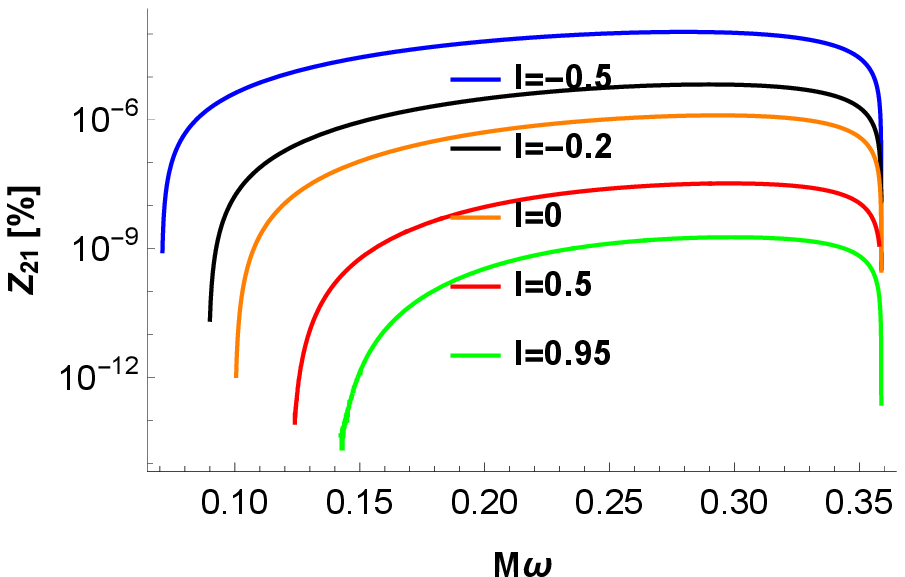}
%\caption{Critical radius for various values of $k$ with $ l=.1,b=.1$ and $\theta=\pi/2$}
\end{subfigure}
\begin{subfigure}{.58\textwidth}
\centering
  \includegraphics[width=.8\linewidth]{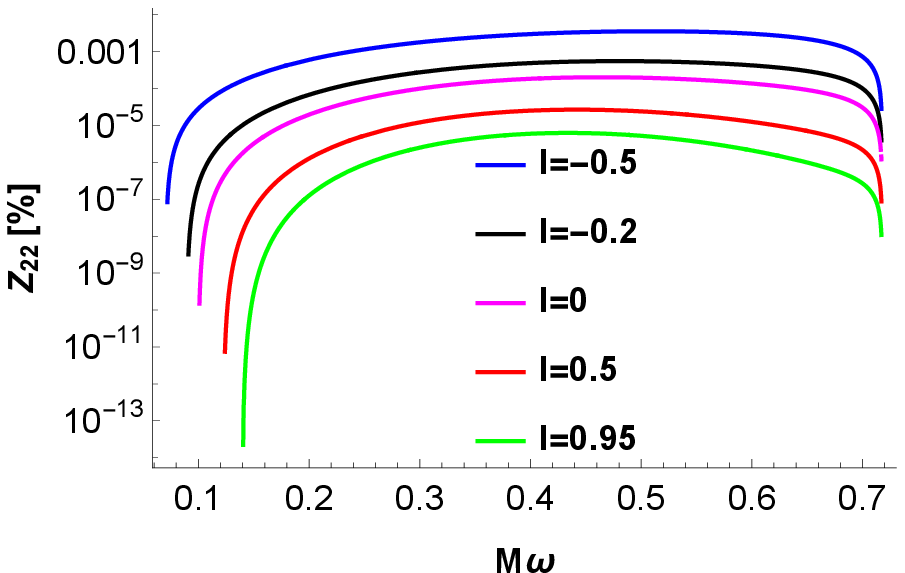}
%\caption{Critical radius for various values of $l$ with $k=.1,b=.1$ and $\theta=\pi/2$}
\end{subfigure}
\caption{Variation of amplification factors with $\ell$ for
various multipoles with $\hat{\mu}=0.1,b=0.1$, and $\hat{a}=0.9$.}
\label{fig:test}
\end{figure}

\begin{figure}[H]
\centering
\begin{subfigure}{.58\textwidth}
%\centering
  \includegraphics[width=.8\linewidth]{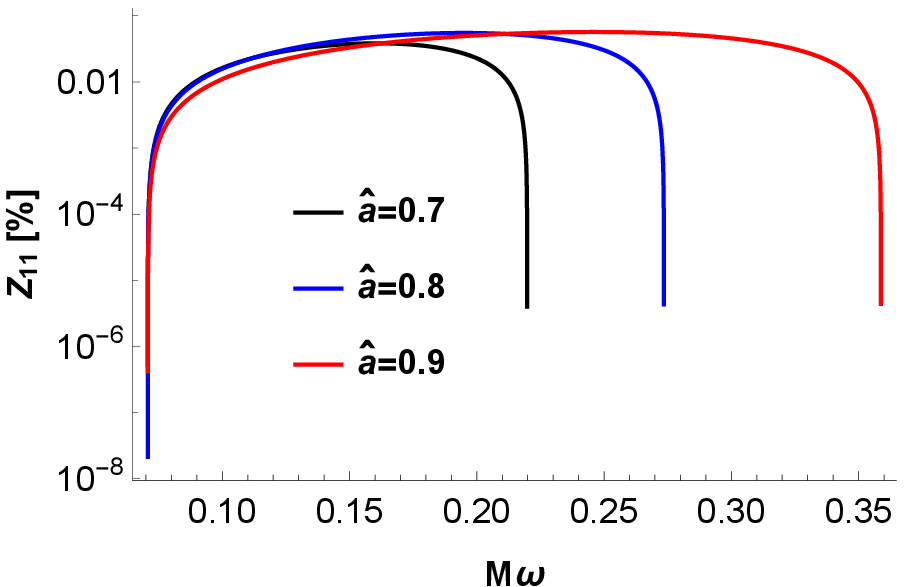}
%\caption{Critical radius for various values of $b$ with $ l=.1,k=.1$ and $\theta=\pi/2$}
\end{subfigure}%
\begin{subfigure}{.58\textwidth}
%\centering
  \includegraphics[width=.8\linewidth]{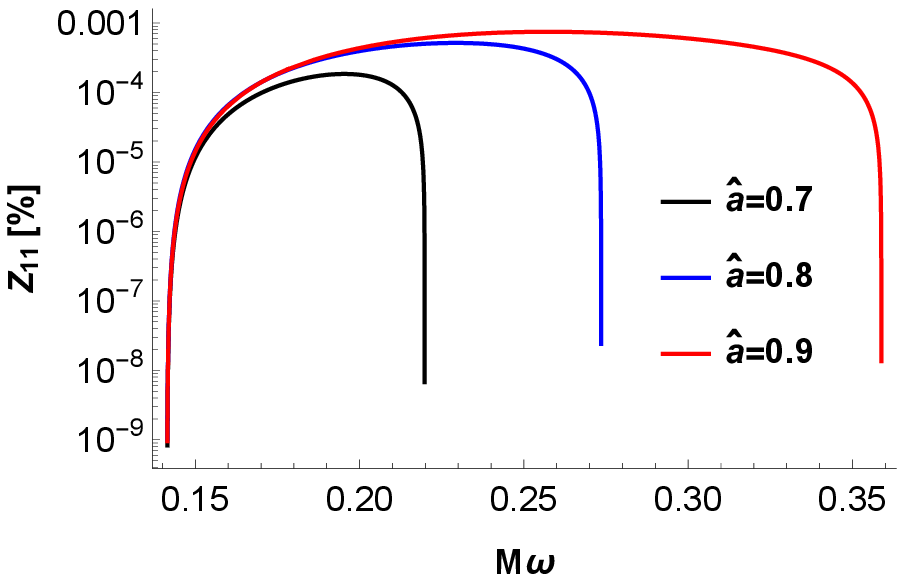}
%\caption{Critical radius for various values of $k$ with $ l=.1,b=.1$ and $\theta=\pi/2$}
\end{subfigure}
\begin{subfigure}{.58\textwidth}
%\centering
  \includegraphics[width=.8\linewidth]{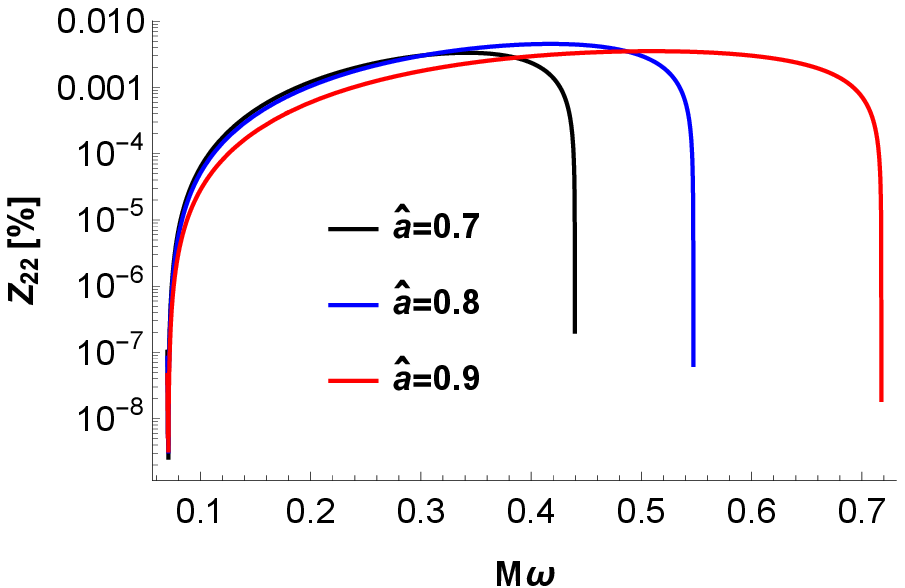}
%\caption{Critical radius for various values of $l$ with $k=.1,b=.1$ and $\theta=\pi/2$}
\end{subfigure}%
\begin{subfigure}{.58\textwidth}
%\centering
  \includegraphics[width=.8\linewidth]{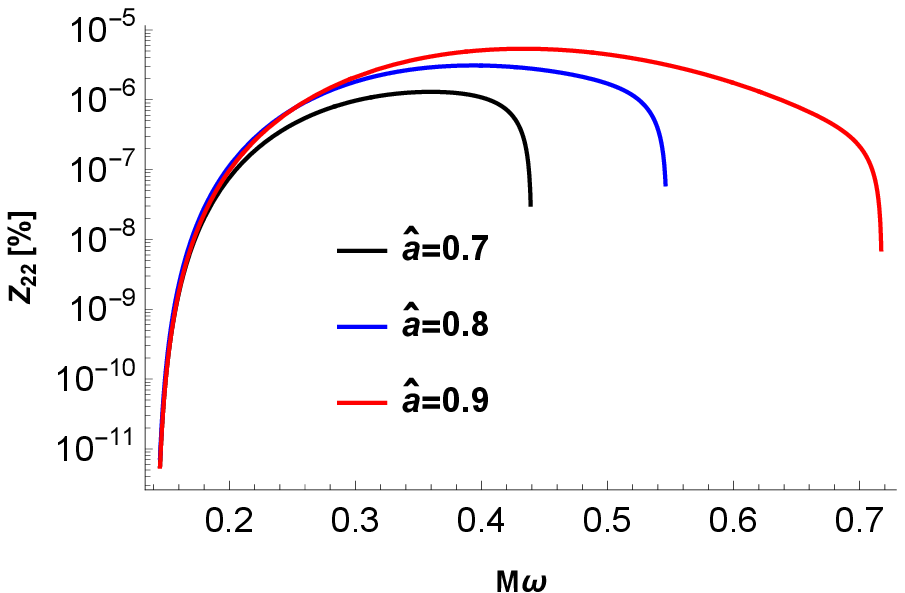}
%\caption{Critical radius for various values of $l$ with $k=.1,b=.1$ and $\theta=\pi/2$}
\end{subfigure}
\caption{Variation of amplification factors with $\hat{a}$ for
various multipoles with $\hat{\mu}=0.1$ and $b=0.1$. For left ones
$\ell=-0.5$ and for right ones $\ell=1$.} \label{fig:test}
\end{figure}

\begin{figure}[H]
\centering
\begin{subfigure}{.58\textwidth}
%\centering
  \includegraphics[width=.8\linewidth]{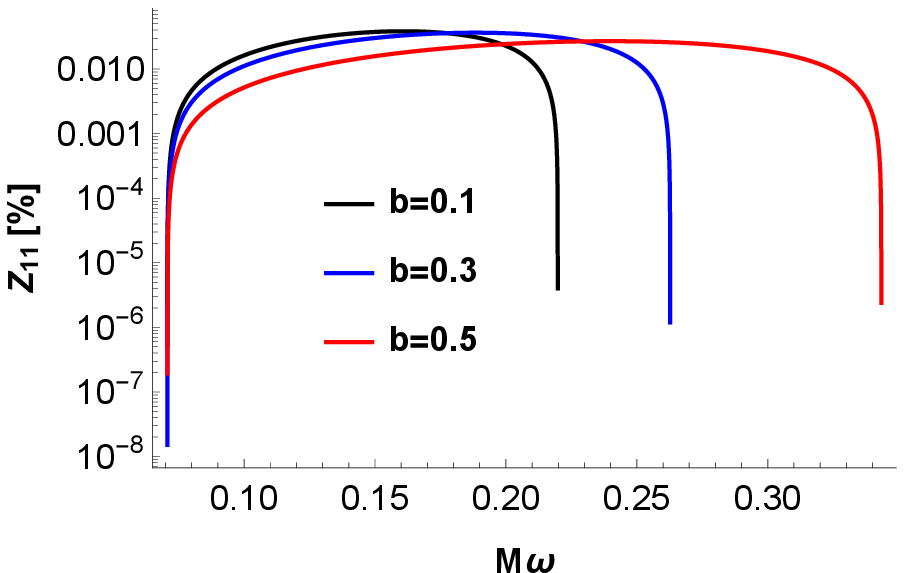}
%\caption{Critical radius for various values of $b$ with $ l=.1,k=.1$ and $\theta=\pi/2$}
\end{subfigure}%
\begin{subfigure}{.58\textwidth}
%\centering
  \includegraphics[width=.8\linewidth]{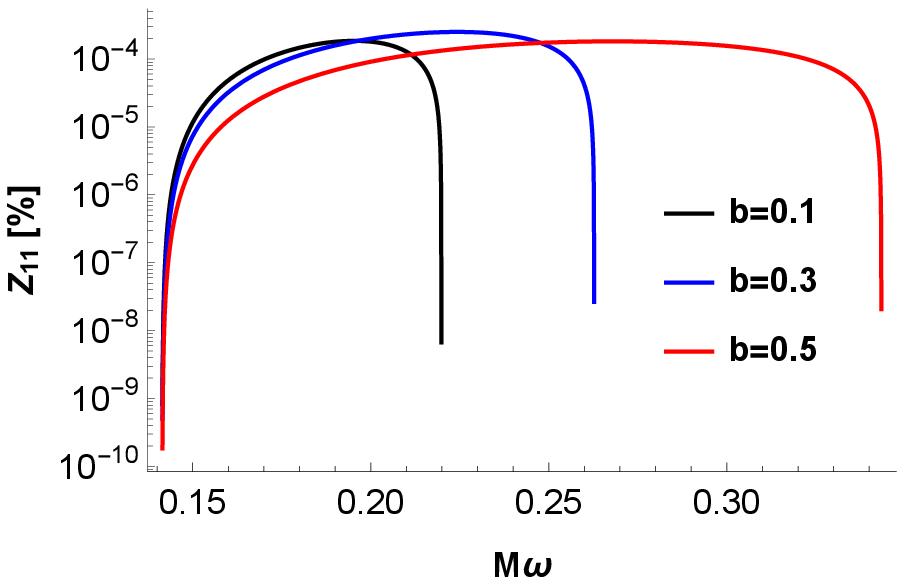}
%\caption{Critical radius for various values of $k$ with $ l=.1,b=.1$ and $\theta=\pi/2$}
\end{subfigure}
\begin{subfigure}{.58\textwidth}
%\centering
  \includegraphics[width=.8\linewidth]{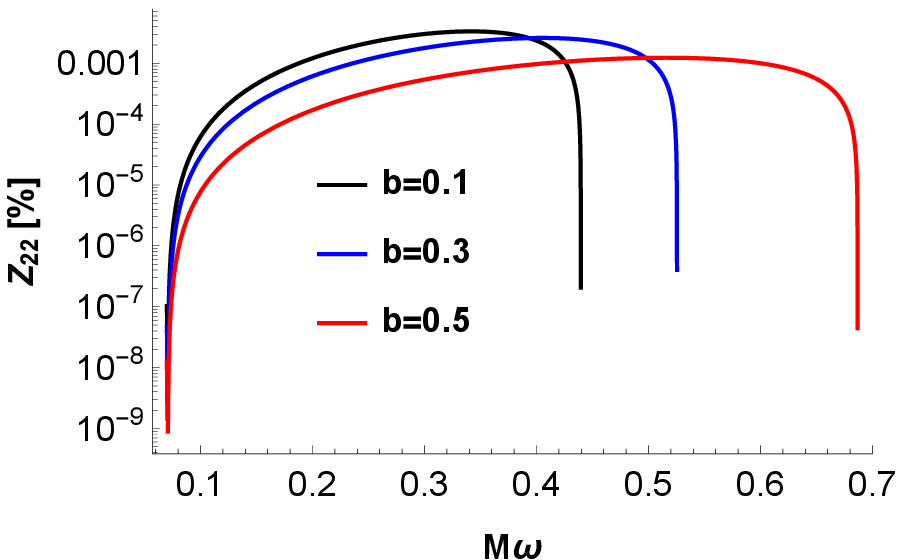}
%\caption{Critical radius for various values of $l$ with $k=.1,b=.1$ and $\theta=\pi/2$}
\end{subfigure}%
\begin{subfigure}{.58\textwidth}
%\centering
  \includegraphics[width=.8\linewidth]{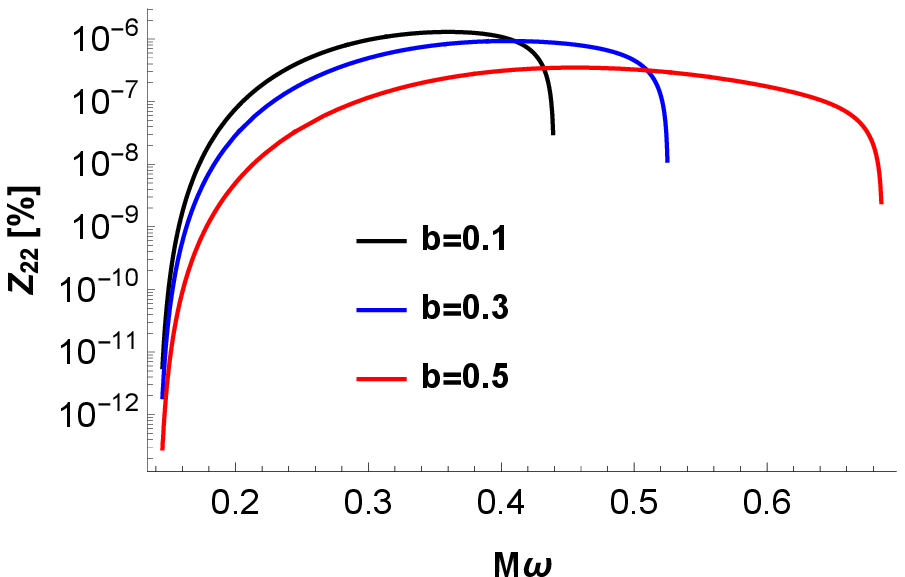}
%\caption{Critical radius for various values of $l$ with $k=.1,b=.1$ and $\theta=\pi/2$}
\end{subfigure}
\caption{Variation of amplification factors with $b$ for various
multipoles with $\hat{\mu}=0.1$ and $\hat{a}=0.7$. For left ones
$\ell=-0.5$ and for right ones $\ell=1$.} \label{fig:test}
\end{figure}
\subsection{Superradiant instability for Kerr-Sen-like black hole}
From equation (\ref{RE}) we have
\begin{eqnarray}
\Delta \frac{d}{d r}\left(\Delta \frac{d R_{\omega j m}}{d
r}\right)+\xi R_{\omega j m}=0, \label{MRE}
\end{eqnarray}
where for a slowly rotating black hole $(\hat{a} \omega \ll 1)$
$$
\xi \equiv\frac{\left(\left(r(r+b)+\hat{a}^{2}\right) \omega-m
\hat{a}\right)^{2}}{1+\ell}+\Delta\left(2 m \hat{a}
\omega-l(l+1)-\mu^{2}r(r+b)\right).
$$
Demanding the black hole bomb mechanism, we should have the
following solutions for the radial equation (\ref{MRE})
$$
R_{\omega j m} \sim\left\{\begin{array}{ll}
e^{-i(\omega-m \hat{\Omega}) r_{*}} & \text { as } r \rightarrow r_{e h}
\left(r_{*} \rightarrow-\infty\right) \\
\frac{e^{-\sqrt{\mu^{2}-\omega^{2} r_{*}}}}{r} & \text { as } r
\rightarrow \infty  \left(r_{*} \rightarrow \infty\right)
\end{array}\right.
$$
The above solution represents the physical boundary conditions
that the scalar wave at the black hole horizon is purely ingoing
while at spatial infinity it is decaying exponentially (bounded)
solution, provided that $\omega^{2}<\mu^{2}$. With the new radial
function
$$
\psi_{\omega j m} \equiv \sqrt{\Delta} R_{\omega j m},
$$
the radial equation (\ref{MRE}) becomes
$$
\left(\frac{d^{2}}{d r^{2}}+\omega^{2}(1+\ell)-V\right)
\psi_{\omega j m}=0.
$$
with
\begin{figure}[H]
\centering
\begin{subfigure}{.58\textwidth}
%\centering
  \includegraphics[width=.8\linewidth]{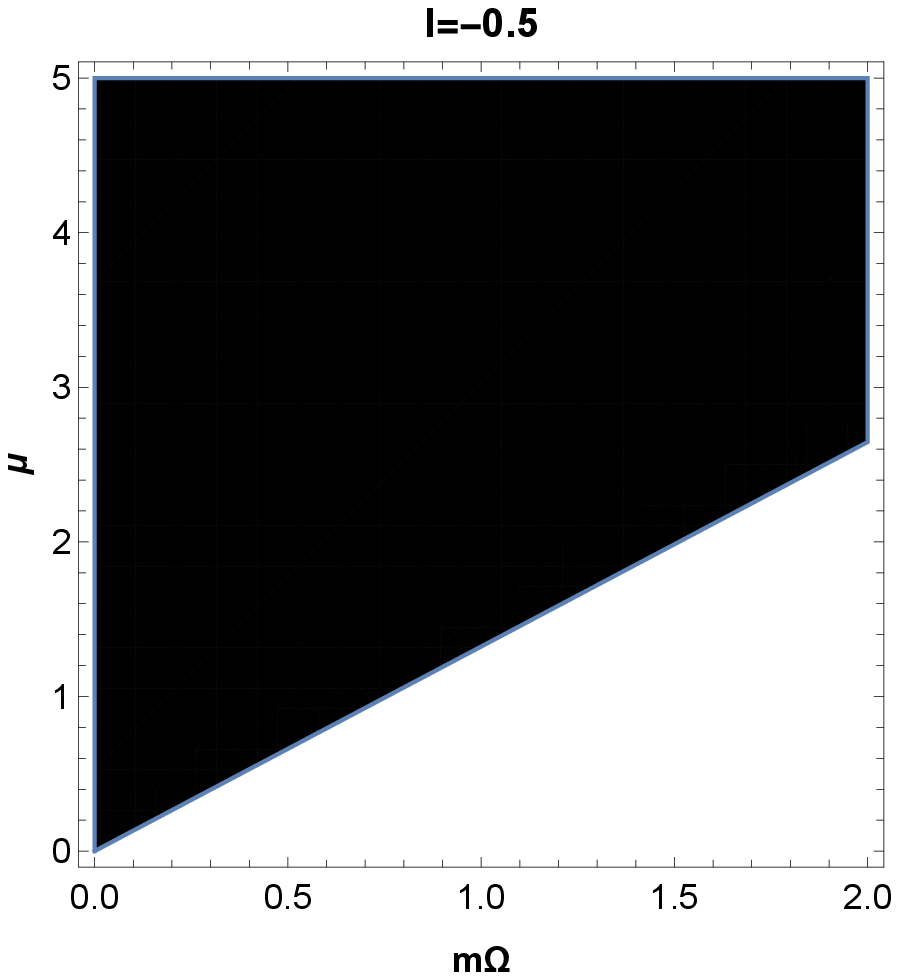}
%\caption{Critical radius for various values of $b$ with $ l=.1,k=.1$ and $\theta=\pi/2$}
\end{subfigure}%
\begin{subfigure}{.58\textwidth}
%\centering
  \includegraphics[width=.8\linewidth]{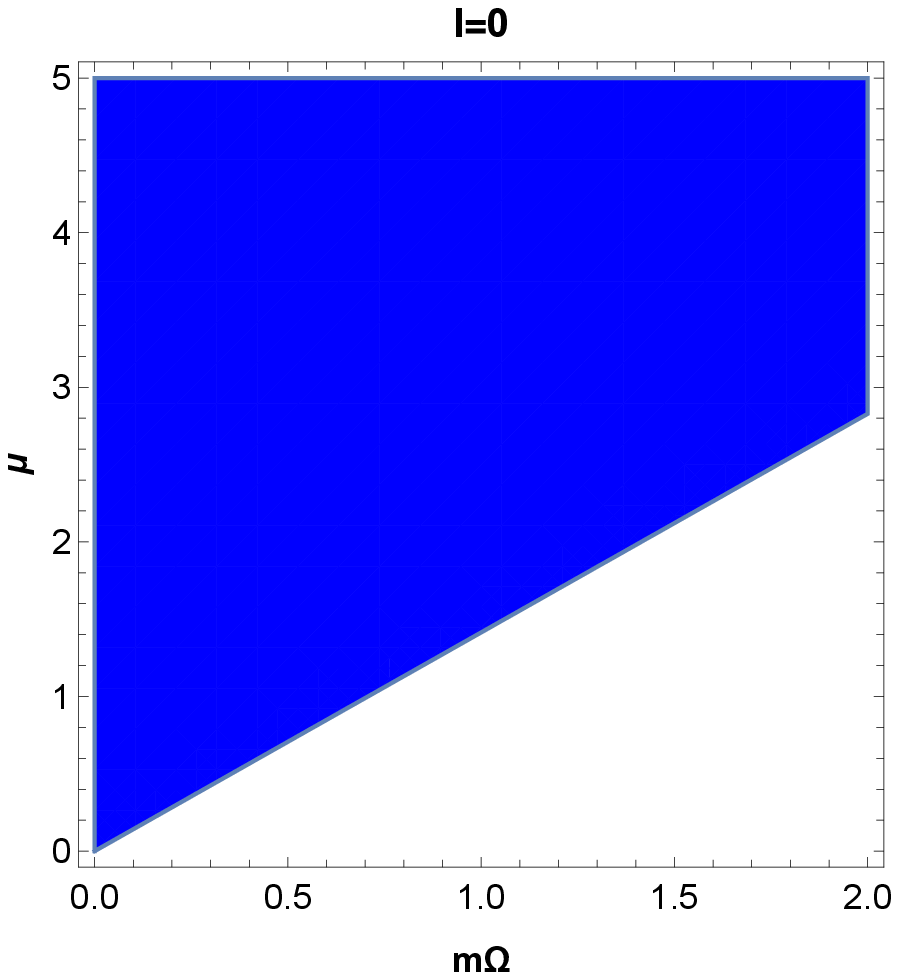}
%\caption{Critical radius for various values of $k$ with $ l=.1,b=.1$ and $\theta=\pi/2$}
\end{subfigure}
\begin{subfigure}{.58\textwidth}
%\centering
  \includegraphics[width=.8\linewidth]{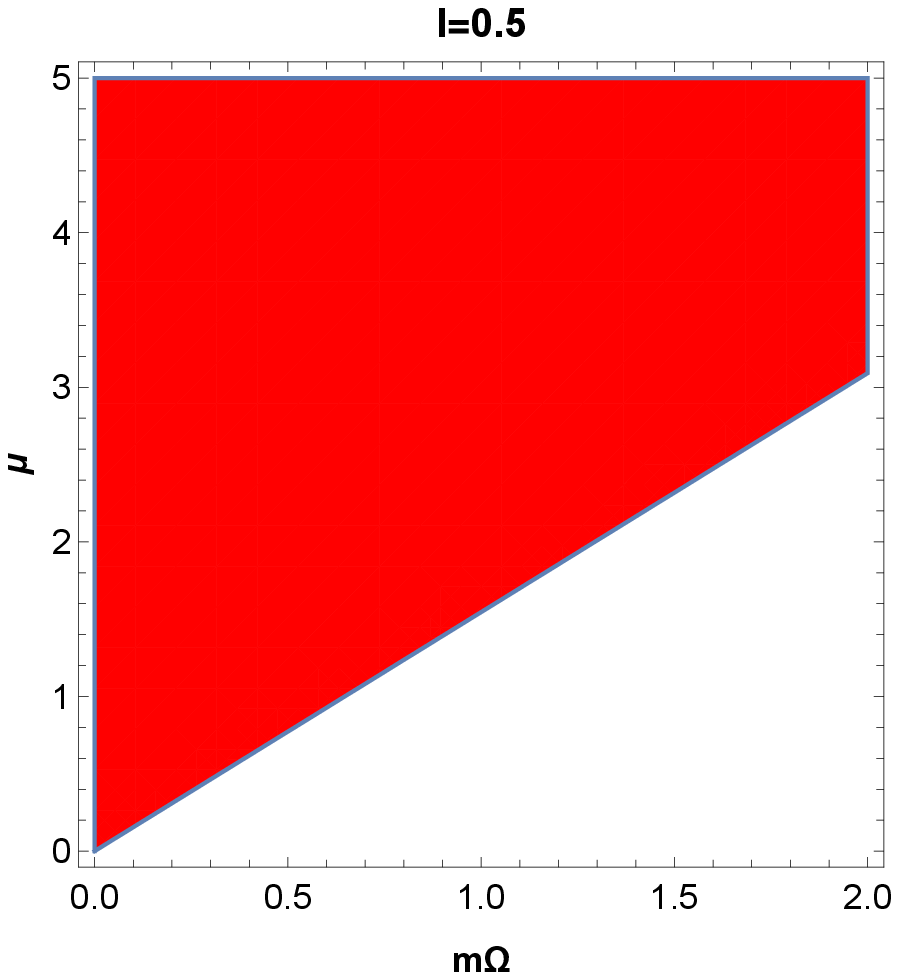}
%\caption{Critical radius for various values of $l$ with $k=.1,b=.1$ and $\theta=\pi/2$}
\end{subfigure}%
\caption{Parameter space($m\Omega$-$\mu$) for massive scalar field
where colored area represents region with stable dynamics and
non-colored area represents region with unstable dynamics.}
\label{fig:test}
\end{figure}
$$
\omega^{2}(1+\ell)-V=\frac{\xi\left(1+\ell\right)+M^{2}-\hat{a}^{2}+\frac{b^{2}}{4}-bM}{\Delta^{2}},
$$
which is the Regge-Wheel equation. By discarding the terms
$\mathcal{O}\left(1 / r^{2}\right)$ the asymptotic form of the
effective potential $V(r)$ reads
$$
V(r)=\mu^{2}\left(1+\ell\right)-\left(1+\ell\right)\frac{4 M
\omega^{2}}{r}+(\ell+1) \frac{2 M \mu^{2}}{r}.
$$
To realize the trapping meaningfully by the above effective
potential it is necessary that its asymptotic derivative be
positive i.e. $V^{\prime} \rightarrow 0^{+}$ as $r \rightarrow
\infty[33]$.  This along with the fact that superradiance
amplification of scattered waves occur when $\omega<m
\hat{\Omega}$ we get the regime
$$
\frac{\mu}{\sqrt{2}}<\omega^{2}<m \hat{\Omega},
$$
in which the integrated system of Kerr-Sen bumblebee black hole
and massive scalar field may experience a superradiant instability,
known as the black hole bomb. The dynamics of the massive
scalar field in Kerr-Sen like black hole will remain stable when
$\mu\geq\sqrt{2}m\hat{\Omega}$.

\section{Constraining from the observed data for $\mathrm{M}87^{*}$}
This section is devoted to constraining the parameters from the
observed data for $\mathrm{M}87^*$. After the announcement of the
capturing of the shadow there have been several attempts to
constraint the parameters used in different modified theories of
gravity \cite{MISBAM87, MK, SGM87}. However, before going towards
constraining parameters let us give a brief description of the
photon orbit in this Kerr-Sen-like spacetime background and see
how the shadow gets deformed with the
additional parameters of this Lorentz violating spacetime.
\subsection{Mathematical formulation of the deviation from circularity}
The Hamiltonian for a massless particle like the photon is given
by
\begin{equation}
H\left(x^{\tau}, p_{\tau}\right)=\frac{1}{2}\left[g^{\tau \nu}
p_{\tau} p_{\nu} -\left(p_{0} \sqrt{-g^{00}}\right)^{2}\right].
\end{equation}
The standard definitions $x^{\tau}=\partial H / \partial p_{\tau}$,
and $\dot{p_{\tau}}=\partial H / \partial x^{\tau}$ renders  the
equations of motion for the photon. Then the null
geodesics in the bumblebee rotating black hole spacetime in terms
of $\xi$ are given by
\begin{eqnarray}\nonumber
\rho^{2} \frac{d r}{d \lambda}=\pm \sqrt{R},\qquad \rho^{2} \frac{d
\theta}{d \lambda}=\pm \sqrt{\Theta}, \\\nonumber
(1+\ell) \Delta \rho^{2} \frac{d t}{d \lambda}=A-2 \sqrt{1+\ell} \operatorname{Mra\xi}, \\
(1+\ell) \Delta \rho^{2} \frac{d \phi}{d \lambda}=2 \sqrt{1+\ell}
M r a+\frac{\xi}{\sin ^{2} \theta}\left(\rho^{2}-2 M r\right),
\label{GEO1}
\end{eqnarray}
where $\lambda$ is the affine parameter and
\begin{eqnarray}
R(r)=\left[\frac{r(r+b)+(1+\ell) a^{2}}{\sqrt{1+\ell}}-a
\xi\right]^{2}-\Delta\left[\eta+(\xi-\sqrt{1+\ell}a)^{2}\right],
\Theta(\theta)=\eta+(1+\ell) a^{2} \cos ^{2} \theta-\xi^{2} \cot
^{2} \theta. \label{GEO2}
\end{eqnarray}
In the Eqns. (\ref{GEO1}) and (\ref{GEO2}), we introduce two
conserved parameters $\xi$ and $\eta$ as usual which are defined
by
\begin{eqnarray}
\xi=\frac{L_{z}}{E} \quad \textrm{and} \quad
\eta=\frac{\mathcal{Q}}{E^{2}},
\end{eqnarray}
where $E$, $L_{z}$, and $\mathcal{Q}$ are the energy, the axial
component of the angular momentum, and the $Carter~constant$
respectively. The radial equation of motion can be cast into the
known form
\begin{eqnarray}
\left(\rho^{2} \frac{d r}{d \lambda}\right)^{2}+V_{e f f}=0.
\end{eqnarray}
The effective potential $V_{e f f}$ in this situation is written
down as
\begin{eqnarray}
V_{e f f}=-[\frac{r(r+b)+(1+l) a^{2}}{\sqrt{1+l}}-a
\xi]^{2}+\Delta\left[\eta+(\xi-\sqrt{1+l}a)^{2}\right]
-\frac{\left[r(r+b)+a^{2}(1+l)\right]^{2}}{1+l}.
\end{eqnarray}
Note that it has explicit dependence on the LV factor $\ell$ and
$a$. So it is natural that the structure the of photon orbit will
depend on the parameters $a$ and $l$. The unstable spherical orbit
on the equatorial plane is given by the following equations
\begin{eqnarray}
\theta=\frac{\pi}{2},\quad R(r)=0,\quad \frac{d R}{d r}=0,\quad \frac{d^{2} R}{d
r^{2}}<0,\quad \mathrm{and}\quad \eta=0.\label{CONDITION1}
\end{eqnarray}
For more generic orbits $\theta \neq \pi / 2$ and $\eta \neq 0,$
the solution of Eqn. (\ref{CONDITION1}) $ r=r_{s}$, gives the $r-$
constant orbit, which is also called spherical orbit and the
conserved parameters of the spherical orbits can be expressed in
the following form
\begin{eqnarray}
\begin{aligned}
\xi_{s} &=\frac{a^{2}\left(1+\mathrm{l}\right)\left(2 M+2
r_{\mathrm{s}} +b\right)+r_{\mathrm{s}}\left(2 r_{\mathrm{s}}^{2}
+3 b r_{\mathrm{s}}+b^{2}-2 M\left(3 r_{\mathrm{s}}
+b\right)\right)}{a\sqrt{1+\mathrm{l}}\left(2 M-2 r_{\mathrm{s}}-b\right)}, \\
\eta_{s} &=-\frac{r_{\mathrm{s}}^{2}\left(-8
a^{2}\left(1+\mathrm{l}\right) M\left(2 r_{\mathrm{s}}+b\right)
+\left(2 r_{\mathrm{s}}^{2}+3b r_{\mathrm{s}} +b^{2} -2 M\left(3
r_{\mathrm{s}}+b\right)\right)^{2}\right)}{a^{2}\left(1+\mathrm{l}\right)\left(2
M-2 r_{\mathrm{s}}-b\right)^{2}} .
\end{aligned}
\end{eqnarray}
The two celestial coordinates which are used to describe the
shape of the shadow that an observer see in the sky,
can be given by
\begin{eqnarray}\nonumber
\alpha(\xi, \eta ; \theta)&=&\lim _{r \rightarrow \infty} \frac{-r
p^{(\varphi)}}{p^{(t)}} = -\xi_{s} \csc \theta,\\\nonumber
\beta(\xi, \eta ; \theta)&=&\lim _{r \rightarrow \infty} \frac{r
p^{(\theta)}}{p^{(t)}}
=\sqrt{\left(\eta_{s}+a^{2} \cos ^{2} \theta-\xi_{s}^{2} \cot ^{2} \theta\right)},\\
\end{eqnarray}
where $(p^{(t)}, p^{(r)}, p^{(\theta)}$, and $p^{(\phi)})$ are the
tetrad components of the photon momentum with respect to locally
non-rotating reference frames \cite{BARDEEN}.
\subsection{Constraining with respect to deviation from circularity data: $\Delta C \le 0.10$}
We now proceed towards constraining the parameters involved in
this spacetime metric from the available experimental findings of
the $M87^*$ as a new window for testing gravity in the
strong-field regime has been opened after the announcement of the
news of capturing the image of supermassive black hole $M87^*$ at
$1.3mm$ wavelength with the angular resolution of $20 \mu as$. The
angular diameter of the shadow of $M87^*$ was fund to be $\theta_d
= 42 \pm 3\mu as$ and the deviation from circularity was $\Delta C
\le 10$ which was consistent with the Kerr black hole's image as
predicted from the theory of General Relativity \cite{EHT1, EHT2,
EHT3, EHT4, EHT5, EHT6}. Let us first proceed to constrain the
parameter from the observation of $M87^*$ concerning $\Delta C \le
.10$. We have considered Kerr-Sen-like black holes, which have
additional parameters $b$, $l$ along with the Kerr black hole
parameters, and the parameters $b$ and $l$ produce deviation from
Kerr geometry with a considerably good configuration. It is also
found that the LV parameter quantitatively influences the
structure of the event horizon by reducing its radius
significantly than that of the Kerr black hole \cite{ARS}, for a
given $b$ and $a$, and the resulting increase in ergosphere area
is thereby likely to have an impact on energy extraction
\cite{ARS}. The  boundary of the shadow is described by the  polar
coordinate $(R(\phi),\phi)$ with the origin at the center of the
shadow $(\alpha_{C}, \beta_{C})$, where
$\alpha_{C}=\frac{|\alpha_{max}+\alpha_{min}|}{2}$ and
$\beta_{C}=0$. If a point $(\alpha, \beta)$ over the boundary of
the image subtends an angle $\phi$ on the $\alpha$ axis at the
geometric center, $\left(\alpha_{C}, 0\right)$ and  $R(\phi)$ be
the distance between the point $(\alpha, \beta)$ and
$\left(\alpha_{C}, 0\right)$, then the average radius
$R_{\text{avg}}$ of the image is given by \cite{CBK}
\begin{equation}
R_{\text {avg}}^{2}  \equiv \frac{1}{2 \pi} \int_{0}^{2 \pi} d \phi R^{2}(\phi), \\
\end{equation}
where $R(\phi) \equiv
\sqrt{\left(\alpha(\phi)-\alpha_{C}\right)^{2}+\beta(\phi)^{2}}$,
and  $\phi = tan^{-1}\frac{\beta(\phi)}{\alpha(\phi)-\alpha_{C}}$

With the above inputs, the  circularity deviation  $\Delta C$ is
defined by \cite{TJDP},
\begin{equation}
\Delta C \equiv 2\sqrt{\frac{1}{2 \pi} \int_{0}^{2 \pi} d
\phi\left(R(\phi)-R_{\text {avg }}\right)^{2}}.
\end{equation}
In the figures below, the deviation
from circularity $\Delta C$ is shown for Kerr-Sen-like black holes
for inclination angles $\theta=90^{o}$ and $17^{o}$ respectively.
\begin{figure}[H]
\centering
\begin{subfigure}{.25\textwidth}
\centering
 \includegraphics[scale=.65]{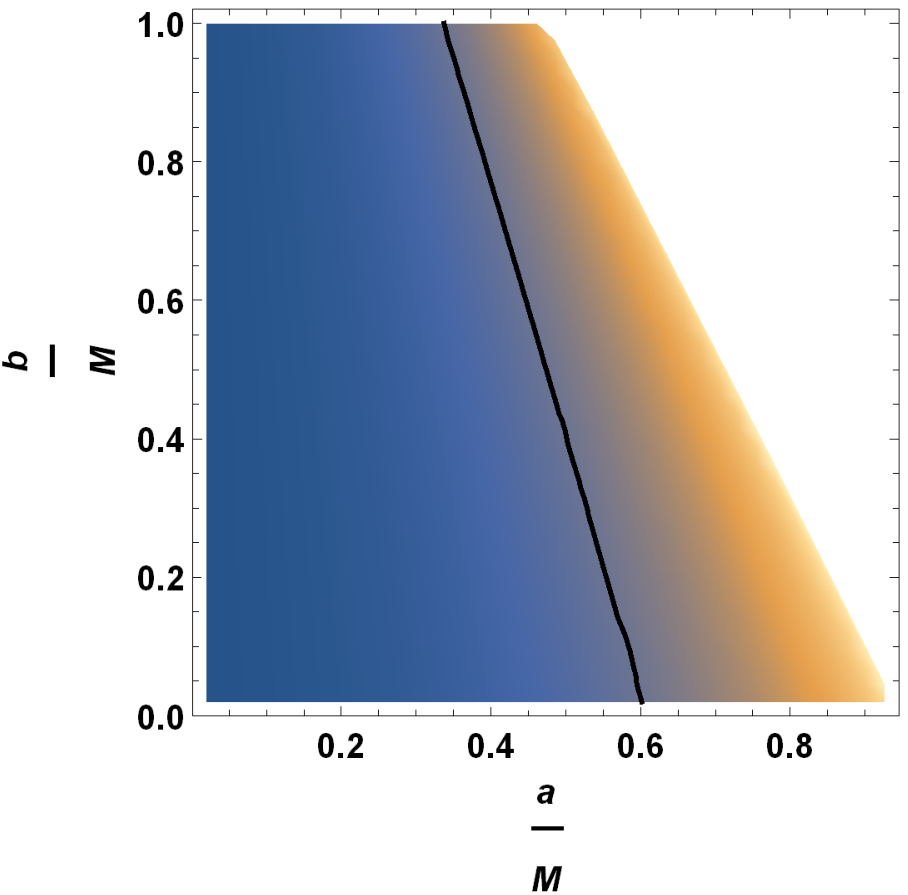}
%\caption{Critical radius for various values of $l$ with $k=.1,b=.1$ and $\theta=\pi/2$}
\end{subfigure}%
\begin{subfigure}{.28\textwidth}
\centering
\raisebox{.2\height}{\includegraphics[scale=.55]{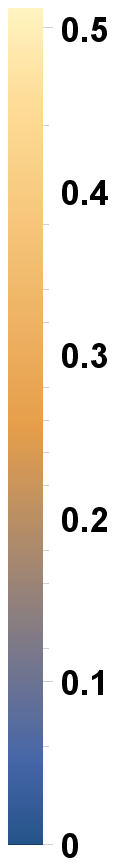}}\hspace{1.5em}%
%\caption{Critical radius for various values of $l$ with $k=.1,b=.1$ and $\theta=\pi/2$}
\end{subfigure}%
\begin{subfigure}{.25\textwidth}
\centering
 \includegraphics[scale=.65]{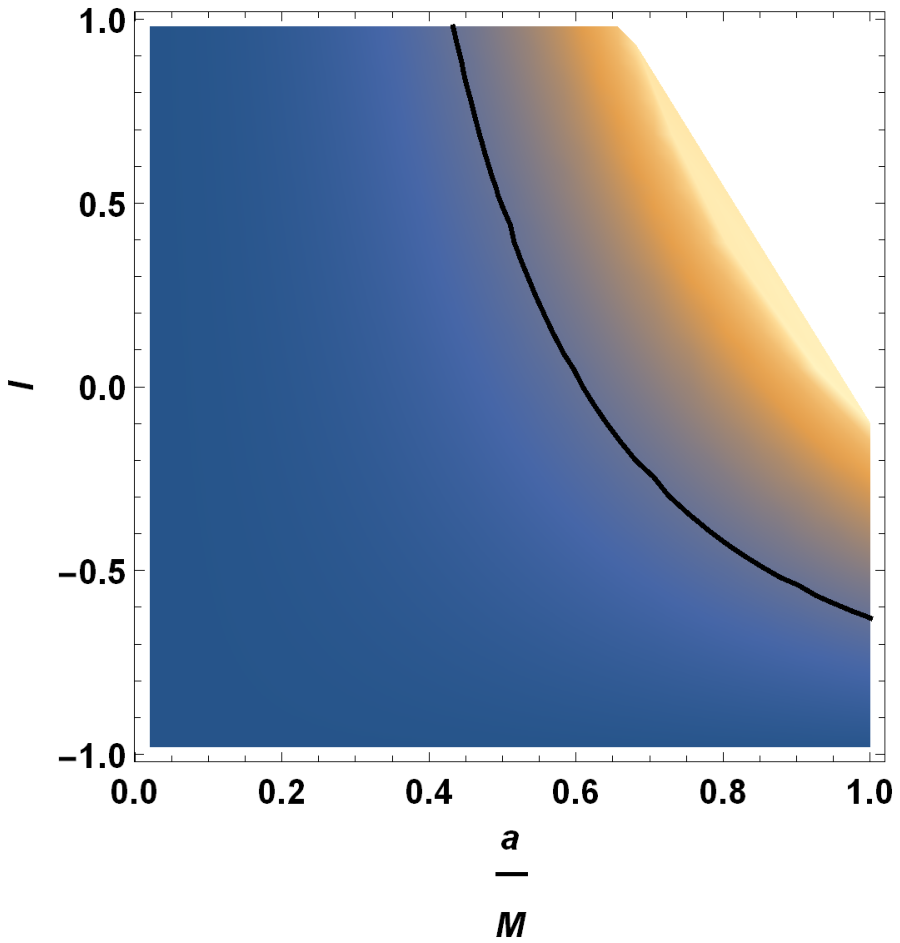}
%\caption{Critical radius for various values of $l$ with $k=.1,b=.1$ and $\theta=\pi/2$}
\end{subfigure}%
\begin{subfigure}{.28\textwidth}
\centering
\raisebox{.19\height}{\includegraphics[scale=.6]{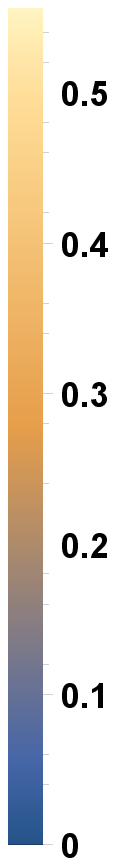}}
%\caption{Critical radius for various values of $l$ with $k=.1,b=.1$ and $\theta=\pi/2$}
\end{subfigure}
\caption{The left one is for $\ell=0.1$ and the right one is for
$b=0.1$ where the inclination angle is $90^{o}$. The black solid
lines correspond to $\Delta C=0.1$.}
%\label{fig:test}
\end{figure}
\smallskip
\begin{figure}[H]
\centering
\begin{subfigure}{.25\textwidth}
\centering
 \includegraphics[scale=.65]{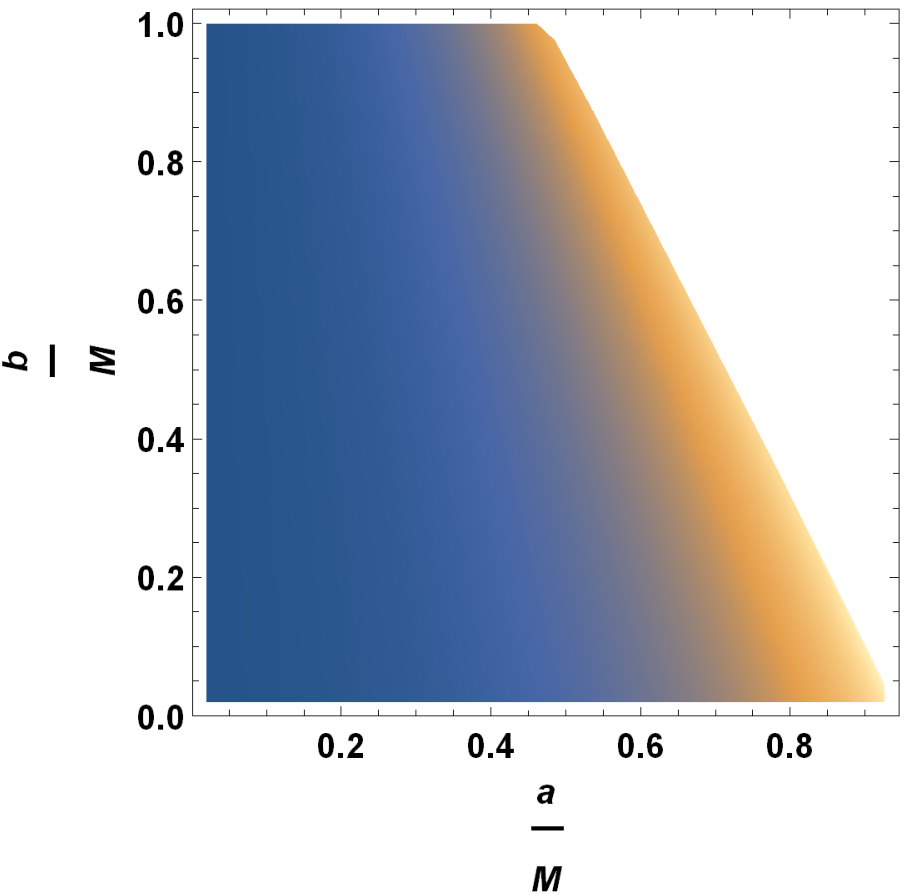}\hspace{1.5em}%
%\caption{Critical radius for various values of $l$ with $k=.1,b=.1$ and $\theta=\pi/2$}
\end{subfigure}%
\begin{subfigure}{.28\textwidth}
\centering
\raisebox{.2\height}{\includegraphics[scale=.55]{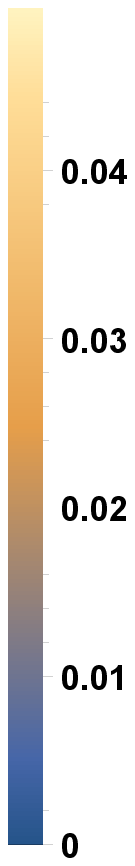}}\hspace{1.5em}%
%\caption{Critical radius for various values of $l$ with $k=.1,b=.1$ and $\theta=\pi/2$}
\end{subfigure}%
\begin{subfigure}{.25\textwidth}
\centering
 \includegraphics[scale=.65]{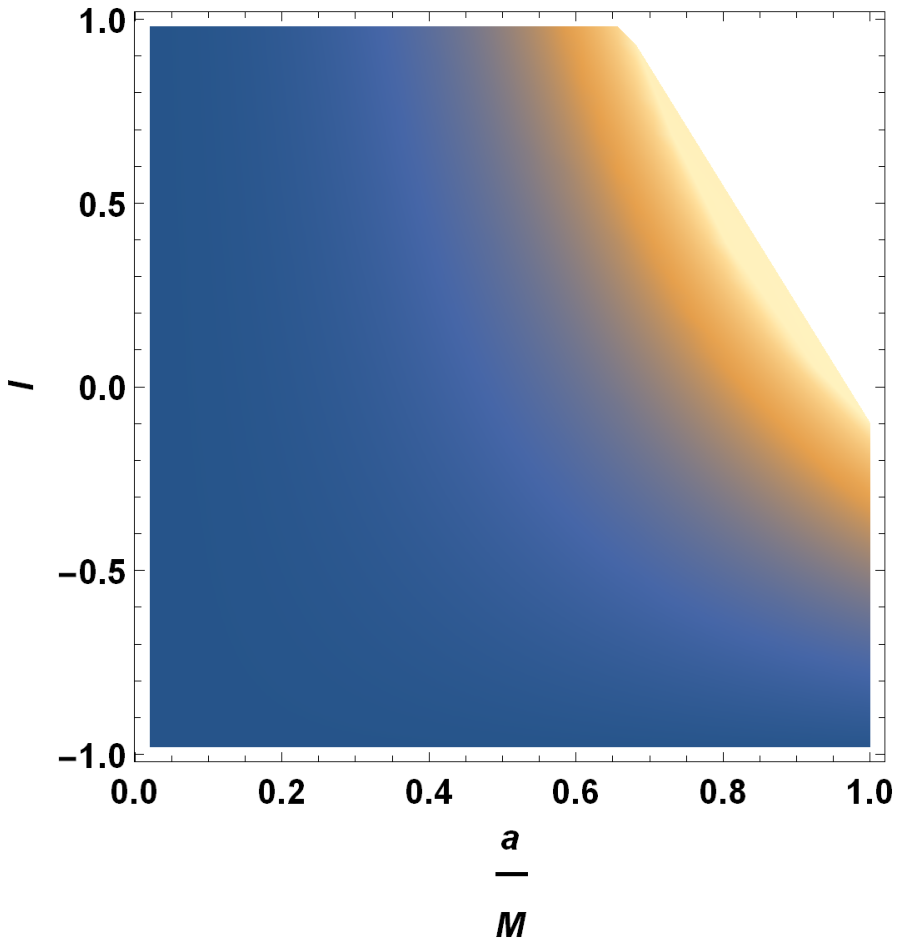}\hspace{1.5em}%
%\caption{Critical radius for various values of $l$ with $k=.1,b=.1$ and $\theta=\pi/2$}
\end{subfigure}%
\begin{subfigure}{.28\textwidth}
\centering
\raisebox{.19\height}{\includegraphics[scale=.6]{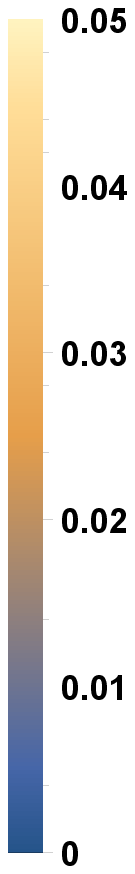}}
%\caption{Critical radius for various values of $l$ with $k=.1,b=.1$ and $\theta=\pi/2$}
\end{subfigure}
\caption{The left one is for $\ell=0.1$ and the right one is for
$b=0.1$ where the inclination angle is $17^{o}$.}
%\label{fig:test}
\end{figure}
We compare the shadows produced from the numerical calculation by
the Kerr-Sen-like black holes with the observed one for the
$\mathrm{M}87^{*}$ black hole. For comparison, we consider the
experimentally obtained astronomical data for the circularity
deviation $\Delta \leq 0.10$ in this subsection. The next section
is devoted to constrain the parameter from observation of angular
diameter $\theta_{d}=42\pm 3 \mu as$ \cite{EHT1, EHT2, EHT3, EHT4,
EHT5, EHT6}.

\subsection{Constraining from the observation of angular diameter $\theta_{d}=42\pm 3 \mu as$}
We now consider the shadow angular diameter which is define by
\begin{equation}
\theta_{d}=\frac{2}{d}\sqrt{\frac{A}{\pi}},
\end{equation}
Where $A=2\int_{r_{-}}^{r_{+}} \beta d\alpha $ is the shadow area
and $d=16.8 Mpc$ is the distance of $M87^{*}$ from the earth.
These relations enable us to accomplish a comparison between the
theoretical predictions for Kerr-Sen-like black-hole shadows and
the experimental findings of the Event Horizon Telescope
collaboration. In the figures below, the angular diameter
$\theta_{d}$ is shown for Kerr-Sen-like black holes for
inclination angles $\theta=90^{o}$ and $17^{o}$ respectively.
\begin{figure}[H]
\centering
\begin{subfigure}{.25\textwidth}
\centering
 \includegraphics[scale=.65]{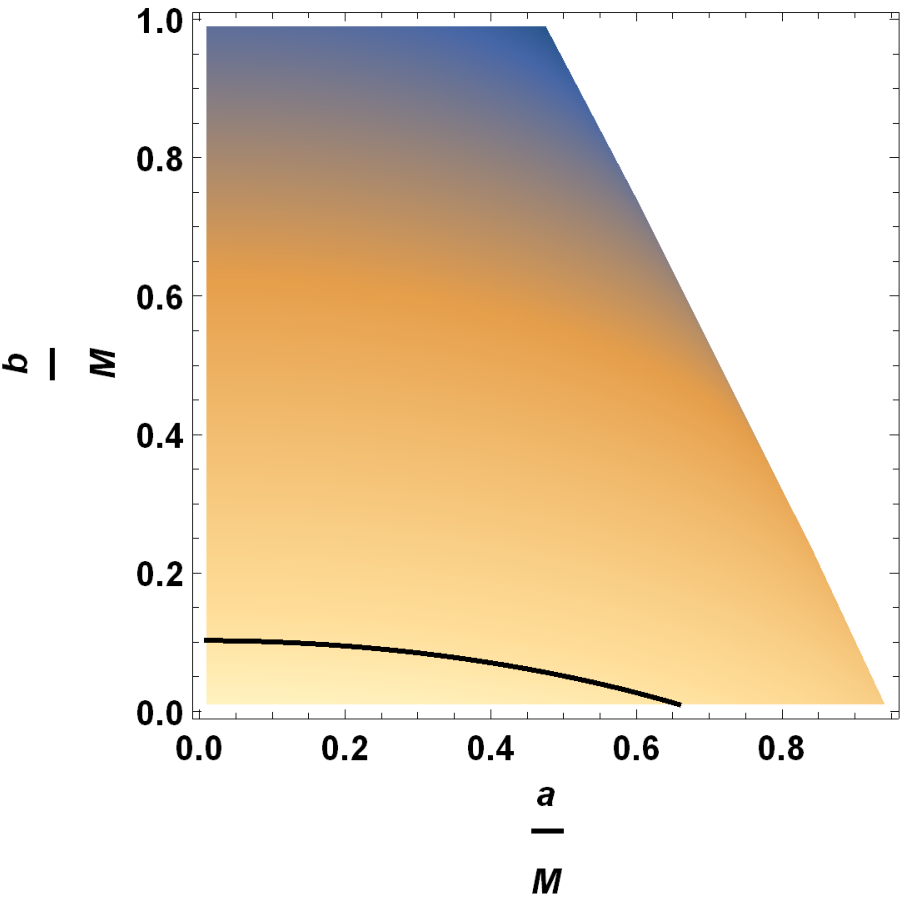}\hspace{1.5em}%
%\caption{Critical radius for various values of $l$ with $k=.1,b=.1$ and $\theta=\pi/2$}
\end{subfigure}%
\begin{subfigure}{.28\textwidth}
\centering
\raisebox{.2\height}{\includegraphics[scale=.55]{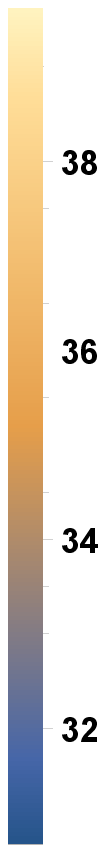}}\hspace{1.5em}%
%\caption{Critical radius for various values of $l$ with $k=.1,b=.1$ and $\theta=\pi/2$}
\end{subfigure}%
\begin{subfigure}{.25\textwidth}
\centering
 \includegraphics[scale=.65]{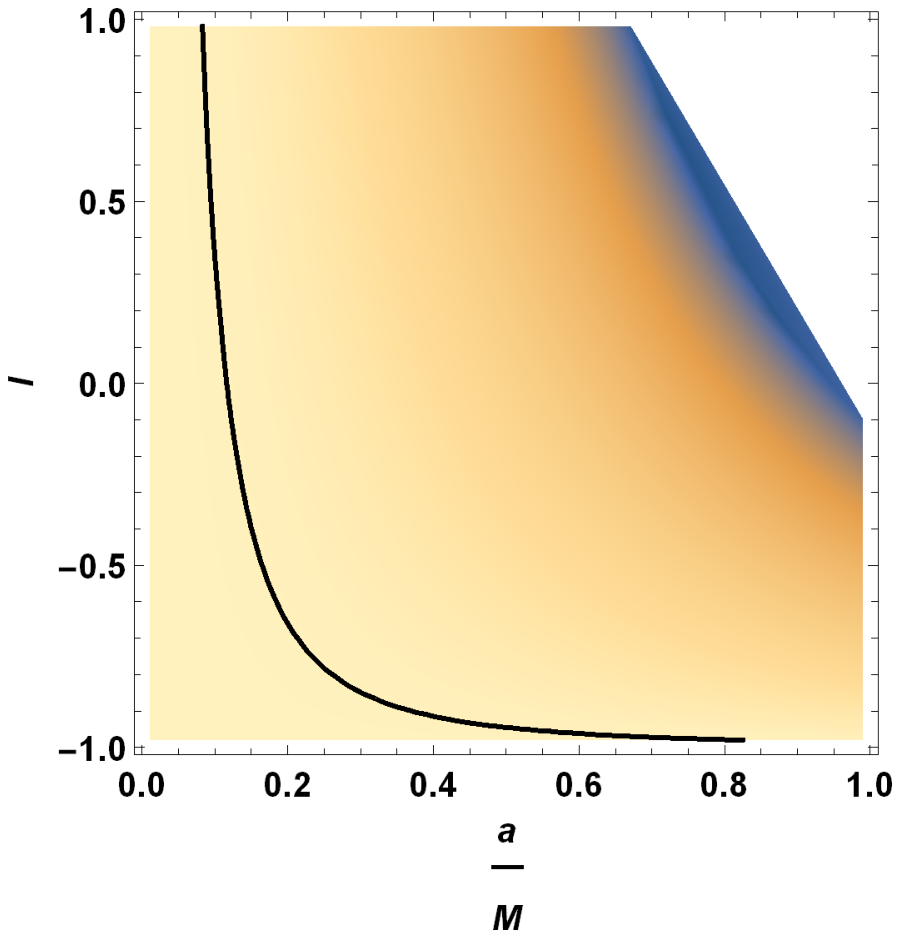}\hspace{1.5em}%
%\caption{Critical radius for various values of $l$ with $k=.1,b=.1$ and $\theta=\pi/2$}
\end{subfigure}%
\begin{subfigure}{.28\textwidth}
\centering
\raisebox{.19\height}{\includegraphics[scale=.6]{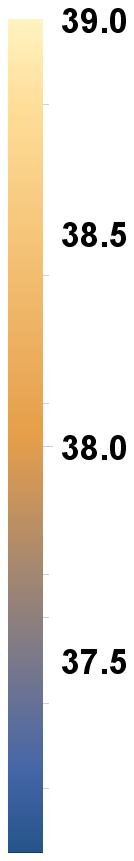}}
%\caption{Critical radius for various values of $l$ with $k=.1,b=.1$ and $\theta=\pi/2$}
\end{subfigure}
\caption{The left one is for $\ell=0.1$ and the right one is for
$b=0.1$ where the inclination angle is $90^{o}$.The black solid lines correspond to $\theta_{d}=39 \mu as$. }
%\label{fig:test}
\end{figure}

\begin{figure}[H]
\centering
\begin{subfigure}{.25\textwidth}
\centering
 \includegraphics[scale=.65]{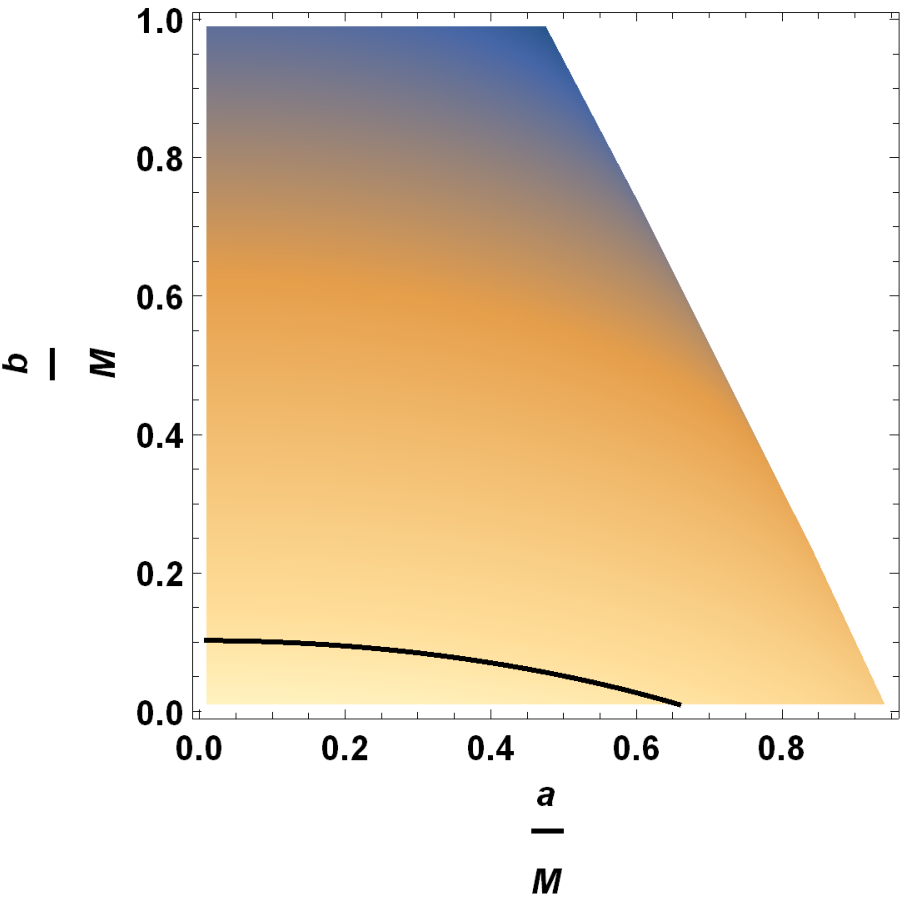}\hspace{1.5em}%
%\caption{Critical radius for various values of $l$ with $k=.1,b=.1$ and $\theta=\pi/2$}
\end{subfigure}%
\begin{subfigure}{.28\textwidth}
\centering
\raisebox{.2\height}{\includegraphics[scale=.55]{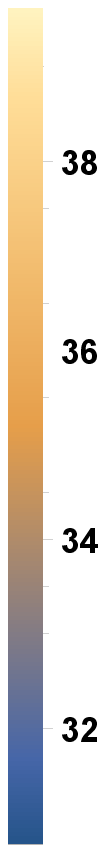}}\hspace{1.5em}%
%\caption{Critical radius for various values of $l$ with $k=.1,b=.1$ and $\theta=\pi/2$}
\end{subfigure}%
\begin{subfigure}{.25\textwidth}
\centering
 \includegraphics[scale=.65]{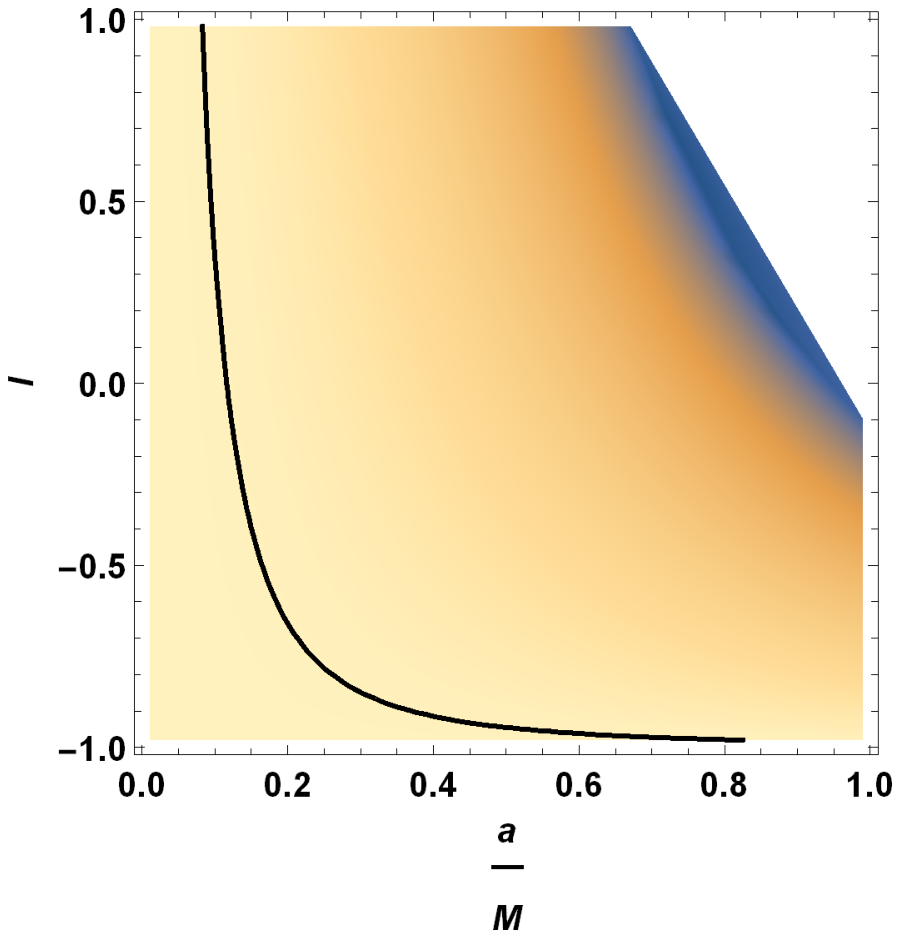}\hspace{1.5em}%
%\caption{Critical radius for various values of $l$ with $k=.1,b=.1$ and $\theta=\pi/2$}
\end{subfigure}%
\begin{subfigure}{.28\textwidth}
\centering
\raisebox{.19\height}{\includegraphics[scale=.6]{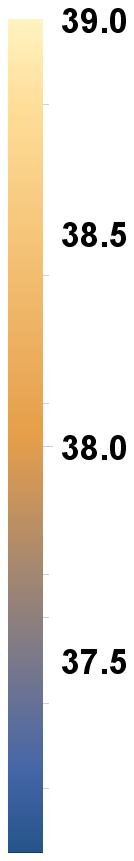}}
%\caption{Critical radius for various values of $l$ with $k=.1,b=.1$ and $\theta=\pi/2$}
\end{subfigure}
\caption{The left one is for $\ell=0.1$ and the right one is for
$b=0.1$ where the inclination angle is $17^{o}$.The black solid lines correspond to $\theta_{d}=39 \mu as$. }
%\label{fig:test}
\end{figure}
\section{Summary and discussions}
In this article, we have studied the superradiance phenomena of
the scalar field scattered off Kerr-sen-like black holes along
with the study of some salient features of the Kerr-sen-like
Lorentz violating spacetime. The LV parameter enters into the
Kerr-Sen-like background via a spontaneous symmetry breaking when
the pseudovector field of the bumblebee field receives a vacuum
expectation value. The Kerr-Sen-like spacetime is a solution to
Einstein's bumblebee gravity model. Along with the parameter $M$,
$a$ (which are contained in the Kerr spacetime metric) the
Kerr-Sen-like metric has two more parameters, $b$ and the LV
parameter $\ell$.The presence of these four parameters has a
crucial role in the spacetime geometry associated with the matric.
Depending on the value of the parameter we can have non-extremal
and extremal cases. We can even have a naked singularity. We
observed that the parameter space formed by $a$ and $b$ that
corresponds to black hole singularity is getting reduced with an
increase in the value of the LV parameter. We have also observed
that the size of the ergosphere gets enhanced with the increase in
the value LV parameter. The increase in the value of $b$ also
results in an increase in the size of the ergosphere like the
increase in the value of $\ell$.

If we look towards the superradiance phenomena of the scalar field
scattered off Kerr-sen-like black holes we find that the LV effect
has a great influence on the superradiance phenomena. How the
superradiance phenomena and the instability associated with it get
influenced by the Lorentz violation effect is studied in detail.
We consider the Klein-Gordon equation in the Kerr-Sen-like
background and employ asymptotic matching of the scalar wave, and
establish transparently from the equation (\ref{AMP}) that in the
low-frequency limit, i.e., for $\omega < m \hat{\Omega}_{e h}$,
the scalar waves show the superradiant mode, i.e. it becomes
amplified in a superradiant manner. The numerical computation,
however, shows that for $m \leq 0$ the massive scalar field has a
non-superradiant mode. For $m \ge 0$ it is superradiant. The role
of the LV parameter in the superradiant phenomena as we have
observed is as follows. The superradiant process enhances with the
decrease in the value of the LV parameter and reverse is the case
when the LV parameter increases irrespective of the sign of the
value of this parameter please vide Fig.6.

Our observation also transpires that the superradiant process gets
influenced by the parameter $b$ also. From Fig. 8, it is clear
that the superradiance enhances with the decrease of the positive
value of the parameter $b$. Note that $b=\frac{Q^2}{M}$. So it
cannot be negative.

Extending the issue of the black hole bomb, the analytical study
of superradiant instability is made. Fig. 9 related to the study
of superradiant instability reveals that the LV parameter
remarkably affects the instability regime. In the background with
negative LV Parameter, the scalar field has more chances to
acquire unstable dynamics and for the positive values of the LV
parameter, these chances are less. Therefore, the LV has a
significant influence on the superradiance scattering phenomena
and the corresponding instability linked with it.

In his article we have also tried to put constraints on parameters
contained in the Kerr-Sen-like spacetime metric from the
observation of $M87^*$, and observed that the circularity
deviation $\delta C \le 0.1$  \cite{EHT1, EHT2, EHT3, EHT4, EHT5,
EHT6} is satisfied exhaustively for the entire $(a-l)$ and $(b-l)$
parameter space at $\theta = 17^0$ inclination angle but at
inclination angle $\theta = 90^0$ it is satisfied for a finite
$(a-l)$ and $(b-l)$ space. The angular diameter satisfies
$\theta_d = 42 \pm 3 \mu as$ within the $1 \sigma$ region
\cite{EHT1, EHT2, EHT3, EHT4, EHT5, EHT6}over a finite $(a-b)$ and
$(a-l)$ space.} However, when we compare with angular diameter
$\theta_d= 42 \pm 3 \mu as$ data it is found that agreeable
$(a-b)$ parameter space is smaller than the $(a-l)$ parameter
space and it is more restricted.

Besides we should make some comments in connection with the recent article \cite{MALUF}.
 The comment of the article although does not target our study of superradiance directly,
 the $b=0$ limit of our metric has to pass through this unfavorable situation. In fact,
  the comment made in \cite{MALUF} is all about the inconsistency of the black
  hole solution obtained in \cite{DING}
which is a special case: b=0 of the metric used here. Even if that
inconsistency is taken as granted in the metric developed in
\cite{DING},  it has been pointed out in the article \cite{KANGI},
that in the slow rotating limit $a\rightarrow 0$ the metric has a
consistent outcome. With this limit,
 the metric turns into a true slowly rotating black hole solution of
 Einstein-bumblebee gravity \cite{DINGC}. Moreover, for $a=b=0$,
 one lands onto the flawless Schwarzschild-like solution of the
 Einstein-bumblebee gravity presented in \cite{CASANA}.

\vspace{.5cm}
 Appendix-A

Following the procedure as followed in \cite{BEZERRA} the general
solution of equation (\ref{RE}) is given by
\begin{equation}
\begin{aligned}
&R(z)=\frac{M}{\sqrt{\Delta(1+1)}} e^{\frac{1}{2} \alpha z}(z-1)^{\frac{1}{2}(1+\gamma)} z^{\frac{1}{2}(1+\beta)} \times\\
&\left\{C_{1}\text {HeunC }(\alpha, \beta, \gamma, \delta, \eta, z)+C_{2} z^{-\beta} \text { HeunC }(\alpha,- \beta, \gamma, \delta, \eta, z)\right\}\\
\end{aligned}
\end{equation}
\text{with}\\
\begin{equation}\nonumber
\begin{aligned}
&2Md=r_{+}-r_{-},z=-\frac{1}{2Md}(r-r_{+})\\
&\alpha=4 d M\left(\mu^{2}-\omega^{2}\right)^{1/2} \sqrt{1+l}\\
&\beta=\sqrt{1-\frac{4 A}{M^{2}}},\\
&\gamma=\sqrt{1-\frac{4 C}{M^{2}}}, \delta=-\frac{2 d}{M^{2}}(B+D)\\
&\eta=\frac{1}{2}+\frac{2Bd}{M^{2}}\\
\end{aligned}
\end{equation} where
\begin{equation}\nonumber
\begin{aligned}
A &=\frac{1}{4 d^{2}}\left[4 M^{4} \omega^{2}(1+l)(d+1)^{2}+\left(M b-\frac{b^{2}}{4}\right) \omega^{2}(1+l)+\hat{a}^{2} m^{2}(1+l)\right.\\
&+4 \hat{a} \omega mMb(1+l)+M^{2} d^{2}+\left(2 M b \omega^{2}\left(Mb-\frac{b^{2}}{4}\right)(1+l)\right.\\
&\left.-4 M^{2}\left(M b-\frac{b^{2}}{4}\right) \omega^{2}(1+l)-4 \hat{a} \omega m M^{2}(1+l)\right)(d+1) \\
&\left.+\left(M^{2} b^{2}-4 M^{3} b\right) \omega^{2}(d+1)^{2}(1+l)\right] \\
\end{aligned}
\end{equation}
\begin{equation}\nonumber
\begin{aligned}
B &=\frac{1}{4 d^{3}}\left[4 M^{4} \omega^{2}(d+1)^{2}(2 d-1)(1+l)+\left(M^{2} b^{2}-4 M^{3} b\right) \omega^{2}\left(d^{2}-1\right)(1+l)\right.\\
&-\left(M b-\frac{b^{2}}{4}\right) \omega^{2}(1+l)-\hat{a}^{2} m^{2}(1+l)-4 \hat{a} \omega mM b(1+l)-M^{2} d^{2} \\
&-\left(2 M b \omega^{2}\left(M b-\frac{b^{2}}{4}\right)(1+l)-4 M^{2}\left(M b-\frac{b^{2}}{4}\right) \omega^{2}(1+l)\right.\\
&-\left.4 \hat{a} \omega m M^{2}(1+l)\right)+2 d^{2}(d+1)\left(-2 M^{3} b \omega^{2}(1+l)\right.\\
&\left.+\mu^{2} M^{3} b(1+l)\right)+2 d^{2}\left(-2 M^{2} \omega^{2}(1+l)\left(M b-\frac{b^{2}}{4}\right)\right.\\
&-\left.\left.\mu^{2} M^{4}(d+1)^{2}(1+l)-\left(\hat{a}^{2}
\omega^{2}+j(1+j)\right) M^{2}(1+l)\right)\right]\\\nonumber
C&=\frac{1}{4 d^{2}}\left[4 M^{4} \omega^{2}(d-1)^{2}(1+l)+\left(M^{2} b^{2}-4 M^{3} b\right) \omega^{2}(d-1)^{2}(1+l)\right. \\
&+\left(M b-\frac{b^{2}}{4}\right) \omega^{2}(1+l)+\hat{a}^{2} m^{2}(1+l)+4 \hat{a} \omega mMb(1+l)+M^{2} d^{2} \\
&-(d-1)\left(2 M b \omega^{2}\left(M b-\frac{b^{2}}{4}\right)(1+l)-4 M^{2}\left(Mb-\frac{b^{2}}{4}\right) \omega^{2}(1+l)\right. \\
&\left.\left.-4 \hat{a} \omega m M^{2}(1+l)\right)\right] \\
D&=\frac{1}{4 d^{3}}\left[4 M^{4} \omega^{2}(d-1)^{2}(2 d+1)(1+l)-\left(d^{2}-1\right)\left(M^{2} b^{2}-4 M^{3} b\right) \omega^{2}(1+l)\right. \\
&+\left(M b-\frac{b^{2}}{4}\right) \omega^{2}(1+l)+\hat{a}^{2} m^{2}(1+l)+4 \hat{a} \omega mMb(1+l)+M^{2} d^{2} \\
&+2 M b \omega^{2}\left(M b-\frac{b^{2}}{4}\right)(1+l)-4 M^{2}\left(M b-\frac{b^{2}}{4}\right) \omega^{2}(1+l) \\
&-4 \hat{a} \omega m M^{2}(1+l)+2 d^{2}(d-1)\left(-2 M^{3} b \omega^{2}(1+l)\right. \\
&\left.+\mu^{2} M^{3} b(1+l)\right)+2 d^{2}\left(2 M^{2} \omega^{2}(1+l)\left(M b-\frac{b^{2}}{4}\right)\right. \\
&\left.\left.+\mu^{2} M^{4}(d+1)^{2}(1+l)+\left(\hat{a}^{2}
\omega^{2}+j(1+j)\right) M^{2}(1+l)\right)-2 \mu^{2}
M^{4}(1+l)\right]
\end{aligned}
\end{equation}
In the limit $z \rightarrow 0 \Rightarrow r \rightarrow r_{+}$ we
have
$$
R(r)=\frac{M}{\sqrt{\Delta(1+1)}}\left[C_{1}\left\{-\frac{1}{2 M
d}\left(r-r_{+}\right)\right\}^{\frac{1}{2}(1+\beta)}+C_{2}\left\{-\frac{1}{2
Md}\left(r-r_{+}\right)\right\}^{\frac{1}{2}(1-\beta)}\right]
\text {,}
$$
and in the limit $z \rightarrow \infty \Rightarrow r \rightarrow
\infty$ we land onto
$$
\begin{aligned}
R(r)&=\frac{M}{\sqrt{\Delta(1+l)}}\left[C_{1} e^{-i \sqrt{\omega^{2}-\mu^{2}} \sqrt{l+l}\left(r-r_{+}\right)} z^{k}\right. \\
&\left.+C_{2} e^{i \sqrt{\omega^{2}-\mu^{2}} \sqrt{1+l}\left(r-r_{+}\right)} z^{-k}\right] \\
\text{where}\\
&k=-\delta / \alpha
\end{aligned}
$$
The solution of near and far region should agree with the solution
used hare in our study of superradence. However it is an involved
mathematical issue and it is beyond the scope of this article.

\end{document}